\documentclass[a4]{elsarticle}
\usepackage{graphicx} 
\usepackage{epsfig} 
\usepackage{float}
\usepackage{subfig}
\usepackage[T1]{fontenc}
\usepackage[english]{babel}
\usepackage[latin1]{inputenc}
\usepackage{amssymb}
\usepackage{epstopdf}
\usepackage{amsmath}
\usepackage{wrapfig}
\usepackage{multicol}
\usepackage{appendix}

\usepackage{color}

\title{Phenomenological model for predicting stationary and non-stationary spectra of wave turbulence in vibrating plates}
\fntext[fn1]{Corresponding author. \\
Email address : thomas.humbert@cea.fr. Tel.: +33 676 01 31 41.}

\address[rvt]{SPEC, CNRS, CEA, Universit� Paris-Saclay, 91191 Gif-sur-Yvette, France}
\address[focal]{Institut d'Alembert, CNRS, UMR 7190, Sorbonne Universit\'es, UPMC, 75005 Paris, France.}
\address[els]{IMSIA, ENSTA ParisTech, CNRS, CEA, EDF, Universit\'e Paris-Saclay, 828 bd des Mar\'echaux, 91762 Palaiseau cedex, France}
 \author[rvt]{T.~Humbert\fnref{fn1}}

 \author[focal]{C.~Josserand}

 \author[els]{C.~Touz\'e}
 \author[els]{O.~Cadot}

\begin{document} 

\maketitle

\subsection*{Abstract}

A phenomenological model describing the time-frequency dependence of the power spectrum of thin plates vibrating in a wave turbulence regime, is introduced. The model equation contains as basic solutions the Rayleigh-Jeans equipartition of energy, as well as the Kolmogorov-Zakharov spectrum of wave turbulence. In the Wave Turbulence Theory framework, the model is used to investigate the self-similar, non-stationary solutions of forced and free turbulent vibrations. Frequency-dependent damping laws can easily be accounted for. Their effects on the characteristics of the stationary spectra of turbulence are then investigated. Thanks to this analysis, self-similar universal solutions are given, relating the power spectrum to both the injected power and the damping law.

\section{Introduction}

The Wave (or Weak) Turbulence Theory (WTT) aims at describing the long-term behaviour of weakly nonlinear systems where the nonlinearity controls the exchanges between scales~\cite{humbert_ref1,humbert_ref2,humbert_ref3}. Under classical assumptions such as dispersivity, weak nonlinearities and the existence of a transparency window in which the dynamics is assumed to be conservative, a kinetic equation can be deduced for the slow dynamics of the spectral amplitude. In addition to the Rayleigh-Jeans spectrum that corresponds to the equipartition of the conserved quantity, here the energy, a broadband Kolmogorov-Zakharov (KZ) spectrum of constant energy flux is predicted, by analogy with hydrodynamic turbulence~\cite{humbert_ref1,humbert_ref2}. Such dynamics has been firstly studied for ocean (gravity) waves~\cite{humbert_ref4,humbert_ref5,humbert_ref6} and since then in systems such as capillary waves~\cite{humbert_ref7,humbert_ref8}, nonlinear optics~\cite{humbert_ref9} or plasmas~\cite{humbert_ref10}.

A wave turbulence spectrum for elastic vibrating plates has been deduced theoretically and observed numerically in~\cite{humbert_ref11}. The theoretical analysis considers the dynamics of a geometrically nonlinear thin vibrating plate in the framework of the F\"{o}ppl-von K{\'a}rm{\'a}n (FVK) equations. 
The WTT analysis leads to the prediction of a direct cascade characterized by a KZ spectrum with constant energy flux.
Soon after, two independent experiments performed on thin elastic plates~\cite{humbert_ref12,humbert_ref13,humbert_ref14} did not recover the theoretically predicted and numerically observed spectra, questioning the validity of the underlying assumptions of WTT in the case of vibrating plates. Recently, an experimental and numerical study considering the effect of damping on the turbulent properties of thin vibrating plates has clearly established that~\cite{humbert_ref15}: 
\begin{itemize}
\item In experiments, damping acts at all scales such that the assumption of a transparency window, a domain in the wave number space where dissipation and injection can be neglected, is questionable.
\item Modifying the damping alters the shape of the velocity power spectra so that a direct comparison with the predicted spectra is out of reach in experimental conditions.
\item However, by including the experimentally measured damping laws in the numerical simulations of the full dynamics (the FVK equations), a good agreement with the experiments is retrieved. This suggests that the discrepancies between the experiments and the WTT predictions are mainly due to damping.
\end{itemize}
These conclusions have been corroborated by a numerical study where the damping was gradually modified, from the experimentally measured law to a vanishing value in a given frequency band \cite{humbert_ref25}, showing also how the spectra are modified by a small yet non-negligible values of damping found in real plates.\\

Accounting for dissipation within the WTT framework remains challenging since the analytic calculations are based on the long time asymptotic evolution of the weakly nonlinear Hamiltonian dynamics. The injection and dissipation in this context can be seen as boundary conditions imposed to the transparency window in the wave number space and to the best of our knowledge, we do not know any analytical attempt to introduce dissipation within the WTT. Another option would be to find an alternative description of the dynamics of the power spectrum, where adding dissipation appears more straightforward. The alternative can be provided by using a phenomenological model describing the temporal evolution of the power spectra, as first proposed by Leith for hydrodynamic isotropic turbulence \cite{humbert_ref16}. These models provide a natural framework for investigating unsteady and self-similar dynamics in  a variety of context ~\cite{humbert_ref16,humbert_ref29,humbert_ref17, humbert_ref18,humbert_ref19,JLTP2006}. They are generally derived from ad-hoc assumptions, by constructing a model equation admitting as stationary solutions both the Rayleigh-Jeans equipartition of energy and the KZ spectrum. This results in a nonlinear diffusion equation in the wave number ($k$-space) or the frequency ($\omega$-space) domain, which mimics the energy transfer within the modes. Thanks to this approach, ideal situations can be investigated, as for instance the injection of a constant flux of energy at small scales and its diffusion, or the evolution of an initial condition in absence of dissipation. Self-similar dynamics are generally observed in these cases.

The goal of this paper is thus to derive and investigate such a phenomenological model in the case of elastic vibrating plates. The model equation should contain both Rayleigh-Jeans and KZ solutions. Injection and dissipation terms are then introduced in order to study more particularly the effects of the damping. Two main results are obtained. First, self-similar dynamics for forced and isolated turbulence in the absence of dissipation are retrieved. In a second part, the effect of the damping on the cascading turbulent spectrum is investigated, exhibiting a self-similar solution relating the power spectrum to the injected power and the damping law.

\section{Model equation}

The application of the wave turbulence theory to the F\"{o}ppl-von K\'arm\'an thin plate equations has been performed in~\cite{humbert_ref11} (see Appendix A for the dimensional and non-dimensional forms of these equations. Note that for this section, all values are dimensionless). Without recalling the details of the derivation and the complex form of the kinetic equation, one only needs to remind that the two stationary solutions of the kinetic equation, written here under the form of a density of energy $E_\omega$, function of the frequency $\omega$, are:
\begin{itemize}
\item The Rayleigh-Jeans equilibrium solution, where the energy $E_\omega$ is equally parted along all the available modes. Consequently, the density of energy $E_\omega$ is a constant that is denoted as $C$:
\begin{equation}
\label{Eq.1b}
E_\omega= C.
\end{equation}
\item The Kolmogorov-Zakharov solution, for which an energy flux $\varepsilon$ is transferred along the cascade until its dissipation near $\omega^{\star}$, the cut-off frequency of the spectrum. Referring to~\cite{humbert_ref11}, the energy spectrum in this case is such that
\begin{equation}
\label{Eq.1}
E^{KZ}_\omega = A\varepsilon^\frac{1}{3} {\rm log}^\frac{1}{3} \left(\frac{\omega^{\star}}{\omega} \right),
\end{equation}
where A is a constant. The specific form of this solution, consisting in a logarithmic correction of the Rayleigh-Jeans spectrum, comes from a degeneracy of the equilibrium solution in a similar manner as for the nonlinear Schr\"odinger equation~\cite{humbert_ref9}. In fact, this logarithmic correction is obtained using a perturbative expansion and is valid far from $\omega^{\star}$. Therefore, although Eq.~\eqref{Eq.1} exhibits a steep cut-off because of the non-existence of the mathematical solution above $\omega^{\star}$ (negative energy), experiments and numerical simulations do not show such a behaviour, and the spectrum decreases more smoothly as $\omega$ increases in the vicinity of $\omega^{\star}$~\cite{humbert_ref15,humbert_ref21,humbert_ref22}.\\
\end{itemize}

The phenomenological model is directly deduced from these stationary solutions of the energy spectrum. Let us consider the following diffusion-like equation in the $\omega$-space for the energy spectrum $E_\omega(\omega,t)$:
\begin{equation}
\partial_t E_\omega= \partial_\omega(\omega E_\omega^2\partial_\omega E_\omega),
\label{Eq.3}
\end{equation}
where $\partial_t$ and $\partial_\omega$ refer respectively for the partial derivatives with respect to time and angular frequency. The energy flux associated to this equation reads straightforwardly
\begin{equation}
\varepsilon = -\omega E_\omega^2 \partial_\omega E_\omega.
\label{Eq.4}
\end{equation}
Thanks to the identification of the energy flux $\varepsilon$, the proportionality constant $A$ of Eq.~\eqref{Eq.1} is then uniquely defined as $A=3^{\frac{1}{3}}$. Hence, for the phenomenological model the KZ solution finally reads : 
\begin{equation}
\label{Eq.Answ}
E^{KZ}_\omega = (3\varepsilon)^\frac{1}{3} {\rm log}^\frac{1}{3} \left(\frac{\omega^{\star}}{\omega} \right).
\end{equation}
The model equation, Eq.~\eqref{Eq.3}, is constructed so that Eq.~\eqref{Eq.1b} and \eqref{Eq.1} are stationary solutions ($\partial_t E_\omega=0$). The Rayleigh-Jeans equilibrium is a trivial solution to Eq.~\eqref{Eq.3} in the stationary case since $\partial_\omega E_\omega=0$. For the KZ spectrum, one has just to verify, by deriving  Eq.~\eqref{Eq.1} with respect to $\omega$, that $\omega E_\omega^2 \partial_{\omega}E_\omega$ is constant with respect to $\omega$. Because this model equation has been deduced in the dimensionless framework, only a numerical prefactor, which could be easily absorbed by a rescaling of the time, should be present on the right-hand side of Eq.~\eqref{Eq.3}. 

The phenomenological equation is nothing else than a nonlinear diffusion equation in the frequency space, in the spirit of the Richardson cascade view of turbulent processes~\cite{humbert_ref20}. 
However, a direct derivation of this equation starting from the kinetic equation cannot be done formally, and only qualitative arguments can be deduced from a local approach on the kinetic equation~\cite{humbert_ref1}(Section 4.3). In fact, attempts to deduce such simplified Fokker-Planck equation from the weak turbulence equations go back to the pioneering works done for ocean waves by Hasselmann~\cite{Hassel1, Hassel2, Hassel3}, although additional approximations were needed to deduce such local models in frequency.

Nonlinear diffusion equations can exhibit important differences as compared to diffusion one. In particular, singularity can be formed by the nonlinear dynamics and compact support solutions can also be present, by opposition to the the linear diffusion where disturbances propagate at infinite speed~\cite{Diaz79}. Here, while a singular cut-off will be observed for the spectra, the equation does not correspond {\it a priori} to the situation were compact support solutions have been proved to exist~\cite{ZelRai}. Finally, it should be said that other phenomenological models exhibiting the same stationary solutions could be deduced and the present model can be considered as one of the simplest among other ones. 

Numerical simulations of this model equation will now be conducted in various cases in order to investigate different dynamical situations. We begin with the classical case where an energy flux is imposed at low frequency and for which the classical KZ spectrum should be observed when dissipation acts at high frequency.

\section{Conservative dynamics of the spectrum in the inertial range}
\subsection{Forced turbulence}
\subsubsection{Non-stationary and stationary spectra}

In order to simulate numerically Eq.~\eqref{Eq.3}, a finite volume method is used. The flux $\varepsilon$ is computed at each frequency increment and the value of $E_\omega$ is defined at the centre of the mesh element. A constant value $\varepsilon_I$ over time for the flux at $\omega=0$ is applied and strong dissipation is included upon $\omega = 10^3$. Remarkably, thanks to this model equation along with this numerical method, simulations exactly corresponding to the ideal configuration of turbulence can be launched, with a flux of energy imposed at $\omega=0$, and dissipation of energy realized with a sink at high frequency. A typical run consists in $2048$ points in the $\omega$ direction, a time step equal to $10^{-7}$ time unit and a total duration of $2$ time units. When the dissipative scale is reached, the cascade front stops its evolution and a stationary regime arises. 

\begin{figure}
\includegraphics[width=6.3cm]{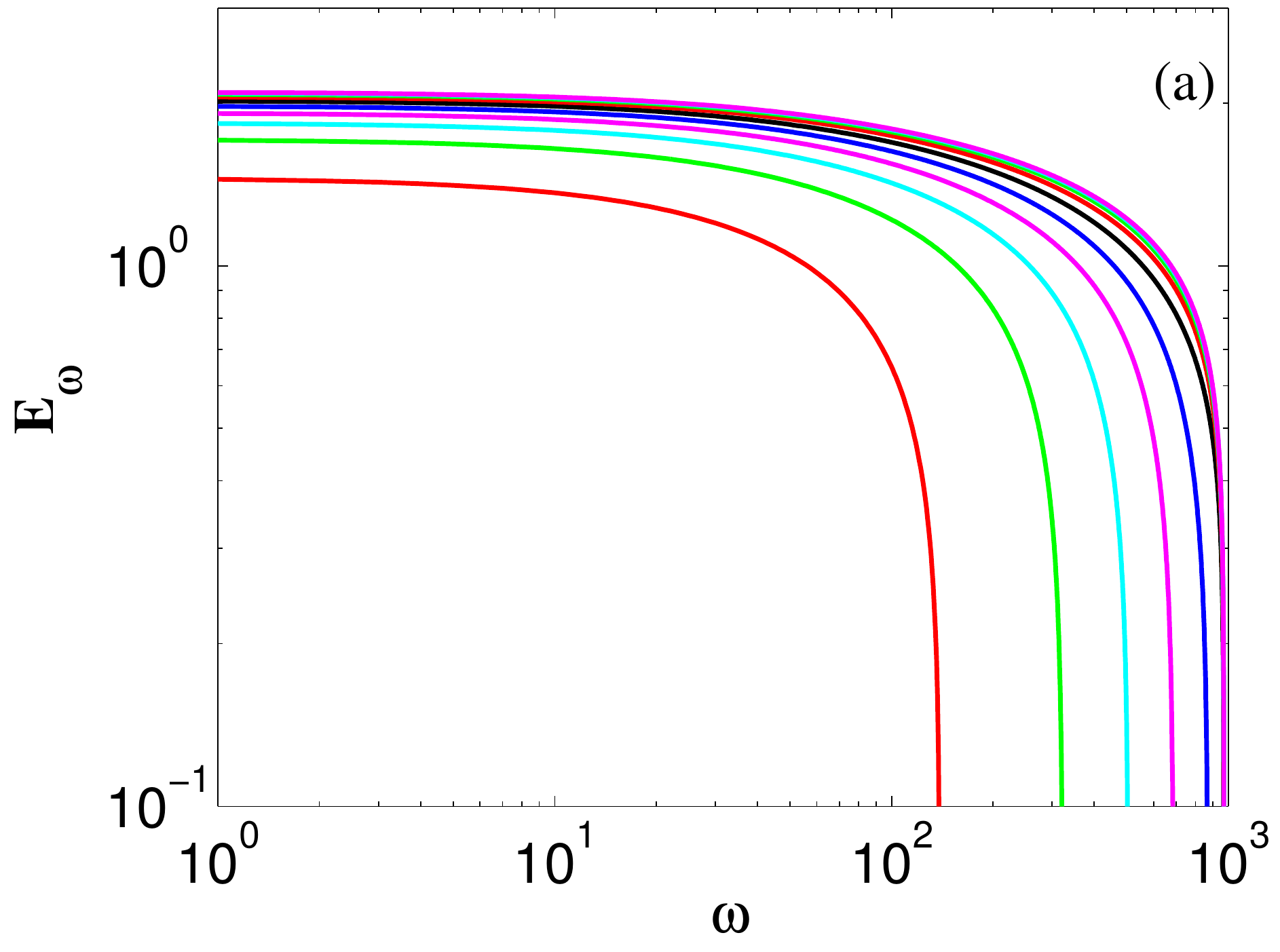}
\includegraphics[width=5.8cm, height=4.8cm]{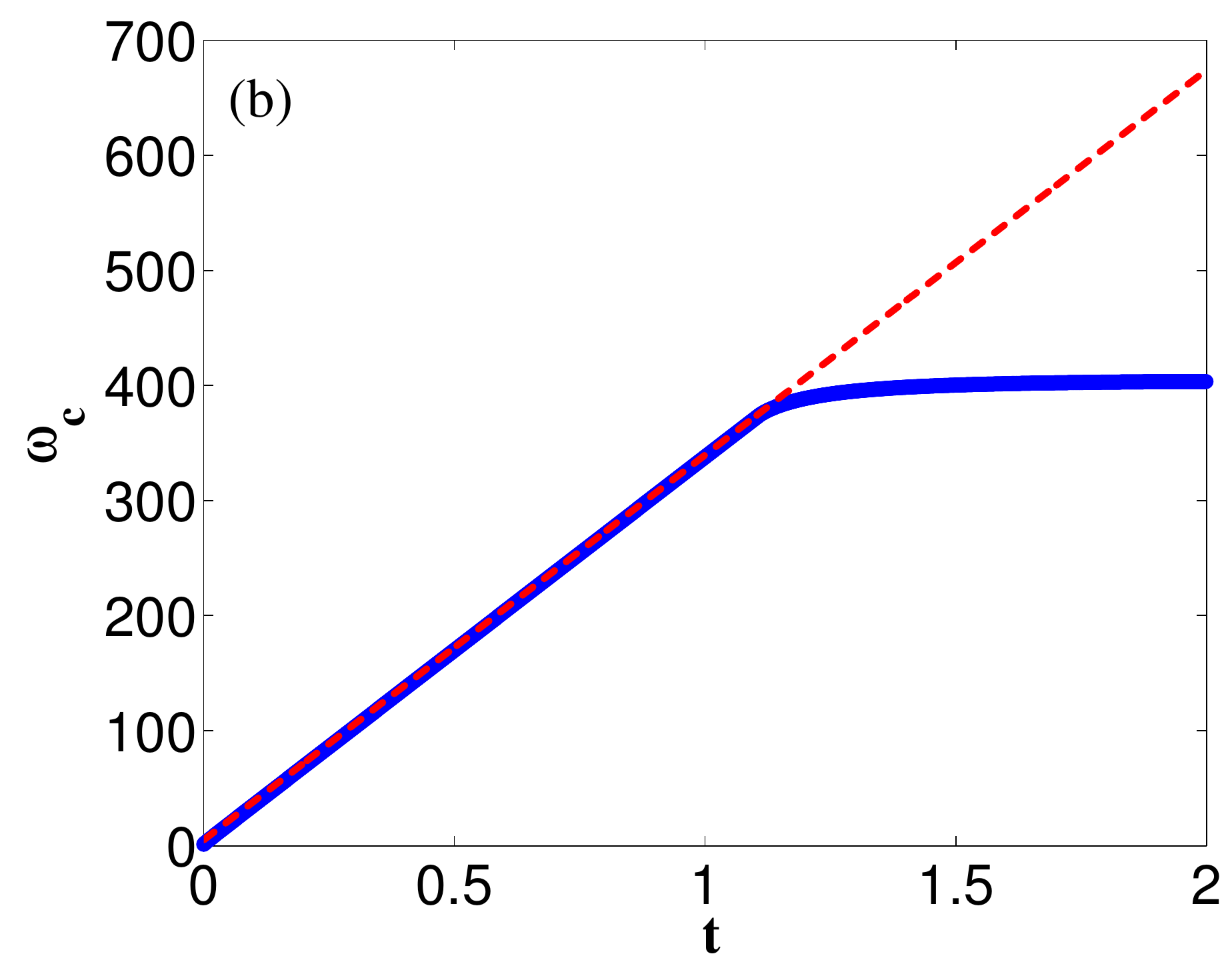}\\
\includegraphics[width=0.5\textwidth]{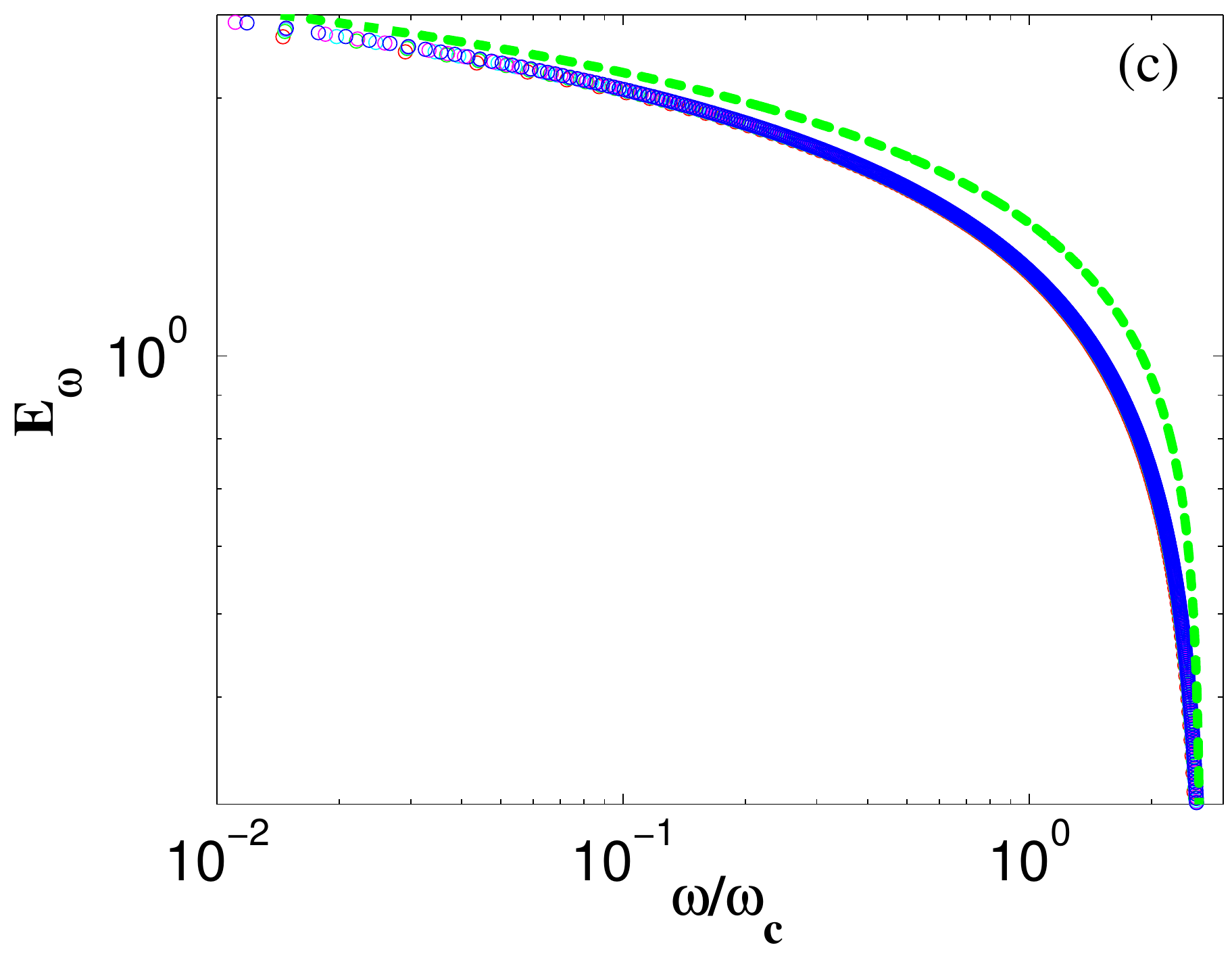}
\includegraphics[width=0.5\textwidth]{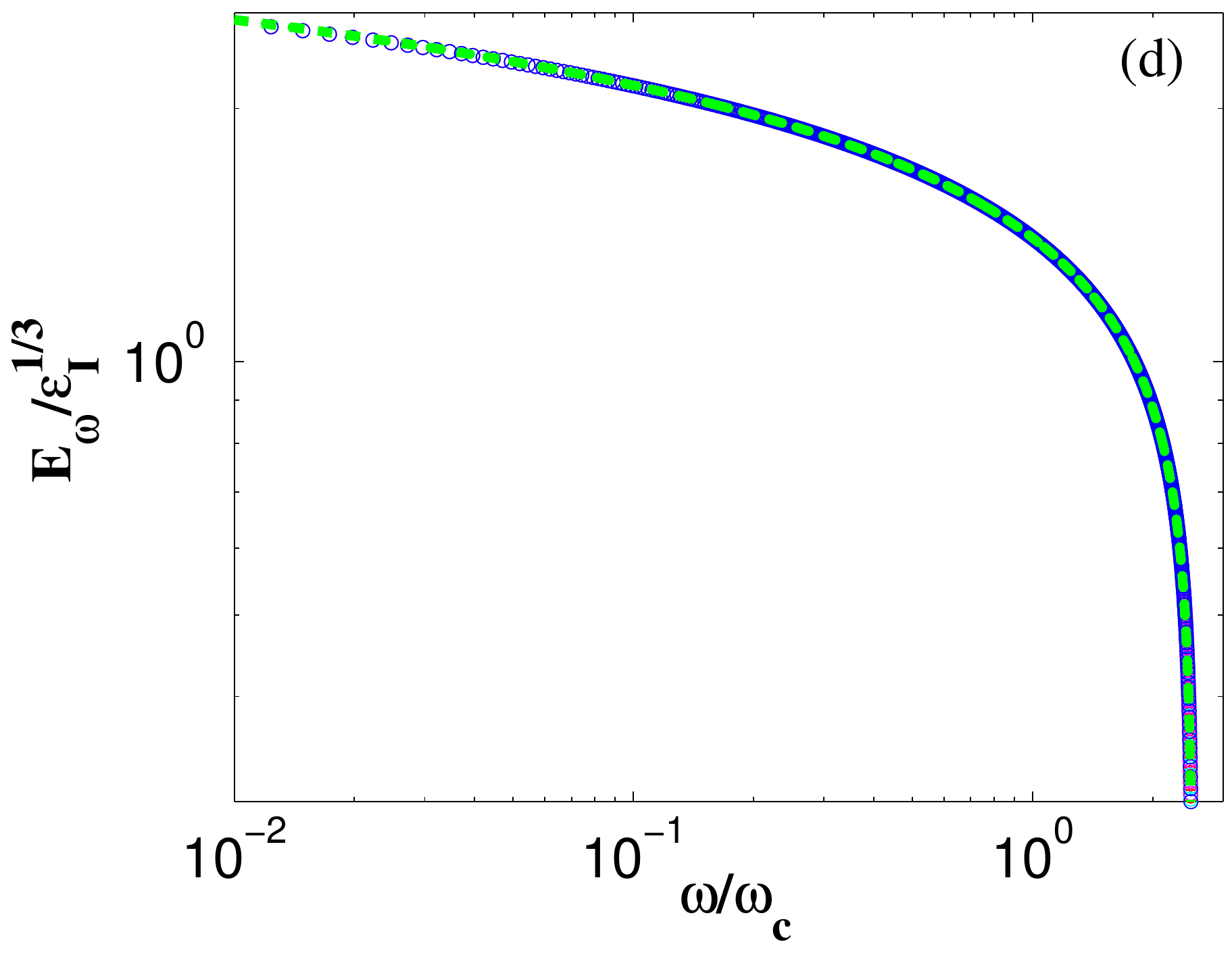}
\caption{\label{Fig1}  Forced turbulence. (a) Energy spectrum $E_\omega$ as a function of the frequency $\omega$, for times increasing from the left to the right, and with $\varepsilon_I = 1$. (b) Characteristic frequency $\omega_c$ defined by Eq.~\eqref{Eq.omega_def} as a function of time. Red dashed line: $\omega_c\propto t$. (c) Energy spectrum $E_\omega(\frac{\omega}{\omega_c})$ computed from the non-stationary spectra (before $t=1$) shown in (a), and compared to the stationary Kolmogorov-Zakharov spectrum $E^{KZ}_\omega = (3\varepsilon_I)^{\frac{1}{3}}\log^{\frac{1}{3}}(\frac{\omega^{\star}}{\omega})$ (green dashed line). (d) Stationary regime. Energy spectrum $E_\omega$, divided by $\varepsilon_I^{1/3}$, plotted as a function of the rescaled frequency $\omega/\omega_c$ for several energy fluxes $\varepsilon_I = 0.5, 1, 2, 5$ and compared to the KZ theoretical spectrum $E^{KZ}_\omega$ (green dashed line).}
\end{figure}

Fig.~\ref{Fig1}(a) displays the energy spectrum every $0.2$ time unit in the considered framework. At the beginning (for $t<1$), the cascade grows towards high frequencies suggesting a self-similar behaviour. More precisely, a characteristic frequency may be  defined as
\begin{equation}
\omega_c=\frac{\int^\infty_{0}E_\omega \omega d\omega}{\int^\infty_{0}E_\omega d\omega},
\label{Eq.omega_def}
\end{equation}
in order to obtain a more quantitative analysis. Fig.~\ref{Fig1}(b) shows the evolution of $\omega_c$ versus time, exhibiting a clear linear behaviour in the transparency window.  When the cascade front reaches the dissipative scale fixed here arbitrarily at $\omega = 10^3$, the characteristic frequency does not evolve anymore and is constant.

Let us first consider the non-stationary regime where the characteristic frequency of the cascade evolves linearly with time for a constant fixed flux. Fig.~\ref{Fig1}(c) displays the non-stationary spectra of Fig.~\ref{Fig1}(a) taken before $t<1$ as functions of the non-dimensional frequency $\omega/\omega_c$. All the curves merge into a unique function, confirming the self-similar growth of the cascade. The shape of this function will be discussed later but can already be compared to the Kolmogorov-Zakharov spectrum Eq.~\eqref{Eq.Answ}, the solution of the phenomenological equation for the conservative case, displayed by a green dashed line in Fig.~\ref{Fig1}(c). Although the two functions are quite close to each other, the self-similar function of the non-stationary regime is steeper near the cut-off. This discrepancy has already been noted in \cite{humbert_ref21}, where the case of forced turbulence within the framework of the F\"{o}ppl-von K\'arm\'an equations (direct simulation) has been studied.

In the stationary regime, shown in Fig.~\ref{Fig1}(d), the phenomenological model recover the Kolmogorov-Zakharov solution for thin plates, as awaited. The scaling of the amplitude of the spectrum by $\varepsilon_I^{1/3}$, as theoretically predicted, is also verified by our data. The typical behaviour of energy spectra of vibrating plates in the stationary regime is therefore correctly described by the phenomenological Eq.~\eqref{Eq.3}. Moreover, our model recovers the fact that the cascade grows with a steeper function of the frequency until its front reaches the dissipative scales, where a stationary regime in agreement with the theoretical predictions arises.

\subsubsection{Self-similar analysis}

In order to recover the numerical behaviour of the non-stationary regime observed in Fig.~\ref{Fig1}(a)(b)(c), the self-similar solutions of Eq.~\eqref{Eq.3} are investigated. The solutions are thus written under the form 
\begin{equation}
E_\omega = t^\alpha g(\frac{\omega}{t^\beta}),
\label{Eq.6}
\end{equation}
with $\alpha$ and $\beta$ two real unknowns and $g$ a function to be determined. Inserting Eq.~\eqref{Eq.6} into Eq.~\eqref{Eq.3}, one finds that $\alpha$ and $\beta$ must fulfil the relationship
\begin{equation}
2\alpha = \beta -1.
\end{equation} 
If we assume further that when injecting with a constant flux over time, the total energy of the plate is growing linearly with time, the equality
\begin{equation}
\int_0^{+\infty}{E_\omega d\omega} = Bt,
\label{Eq.8}
\end{equation}
where $B$ is a constant, leads to a second relationship $\alpha + \beta = 1$. This yields $\alpha = 0$ and $\beta = 1$ so that finally the self-similar solutions are necessarily under the form 
\begin{equation}
E _\omega= g(\frac{\omega}{t}).
\label{Eq.9}
\end{equation}
The previous observation that the characteristic frequency of the self-similar solutions of Eq.~\eqref{Eq.3} in case of forced turbulence grows linearly with time is retrieved. 

Inserting Eq.~\eqref{Eq.9} into Eq.~\eqref{Eq.3}, the equation for the self-similar function $g_\eta=g(\omega/t)$ finally reads
\begin{equation}
-\eta g_\eta' = (\eta g_\eta^2g'_\eta)',
\label{Eq.10}
\end{equation}
where $'$ stands for the derivative with respect to the self-similar variable $\eta=\omega/t$. This equation is solved using Matlab algorithm \textit{ode45} which applies a fourth-order Runge-Kutta scheme with a variable time step~\cite{humbert_ref23}. For this purpose, Eq.~\eqref{Eq.10} is written at the first order:
\begin{equation}
Y' = \begin{pmatrix}
1 & 0\\
0 & -\frac{1}{g^2_\eta}-\frac{1}{\eta}-2\frac{g'_\eta}{g_\eta}
\end{pmatrix}Y, \qquad \mbox{with}\qquad Y =\begin{bmatrix}
g_\eta \\
g'_\eta
\end{bmatrix}\;\mbox{and}\;Y' =\begin{bmatrix}
g'_\eta \\
g''_\eta
\end{bmatrix}.\label{probleme_valeur_init}
\end{equation} 
The initial value problem consists in choosing, for $\eta_0$ given and small (in the simulations, $\eta_0=0.01$ is selected), the values of $g_\eta$ and $g'_\eta$ that determine the desired initial flux $\varepsilon_I$. Whereas the value of $g_\eta(\eta_0)$ is selected for comparison with a given dataset, $g'(\eta_0)$ is retrieved from Eq.~\eqref{Eq.4}. As the flux $\varepsilon_I$ is fixed, one obtains $g'_\eta(\eta_0)=-\frac{\varepsilon_I}{\eta_0 g^2_\eta(\eta_0)}$.

\begin{figure}
\begin{center}\includegraphics[width=0.45\textwidth]{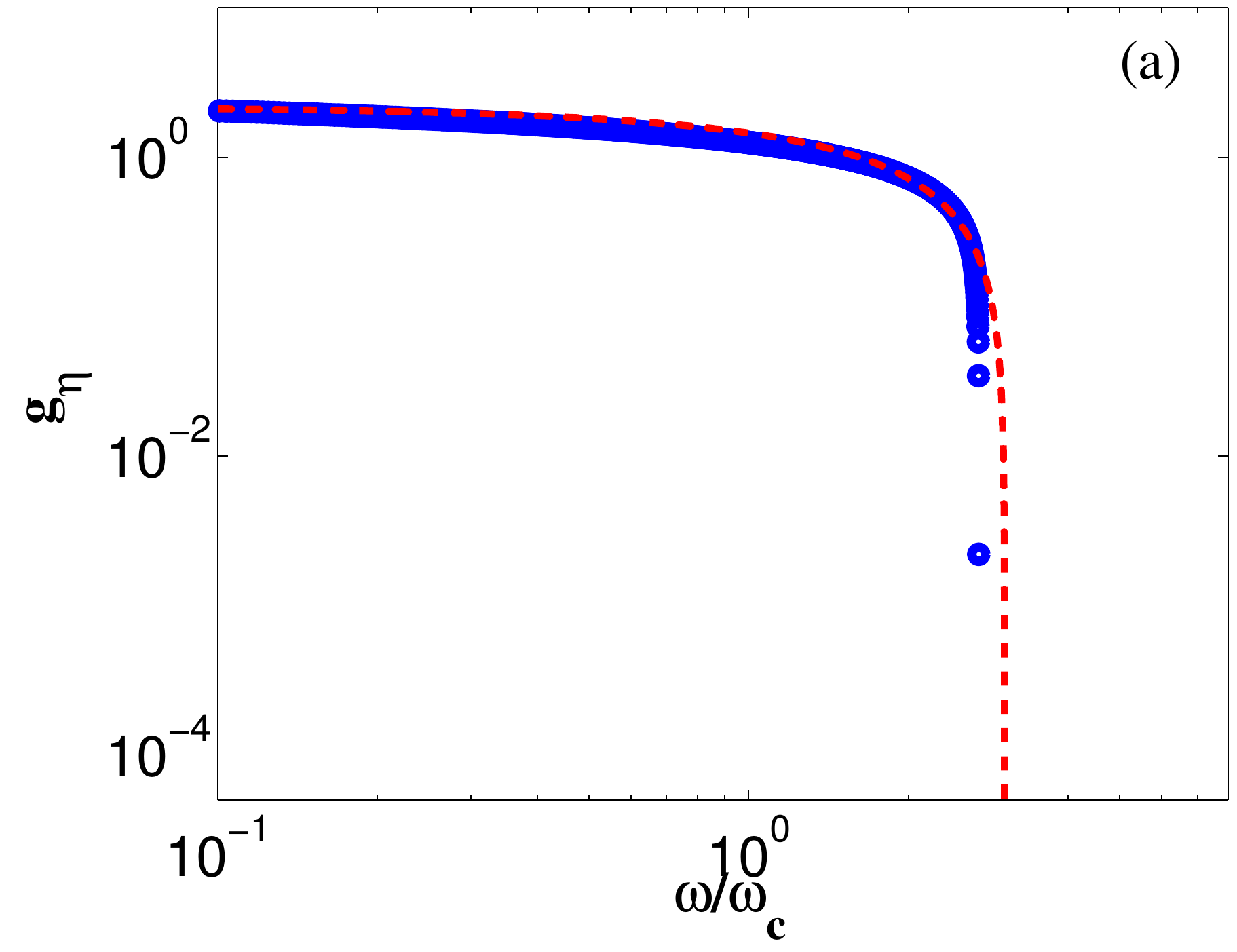}
\hspace{0.5cm}
\includegraphics[width=0.45\textwidth]{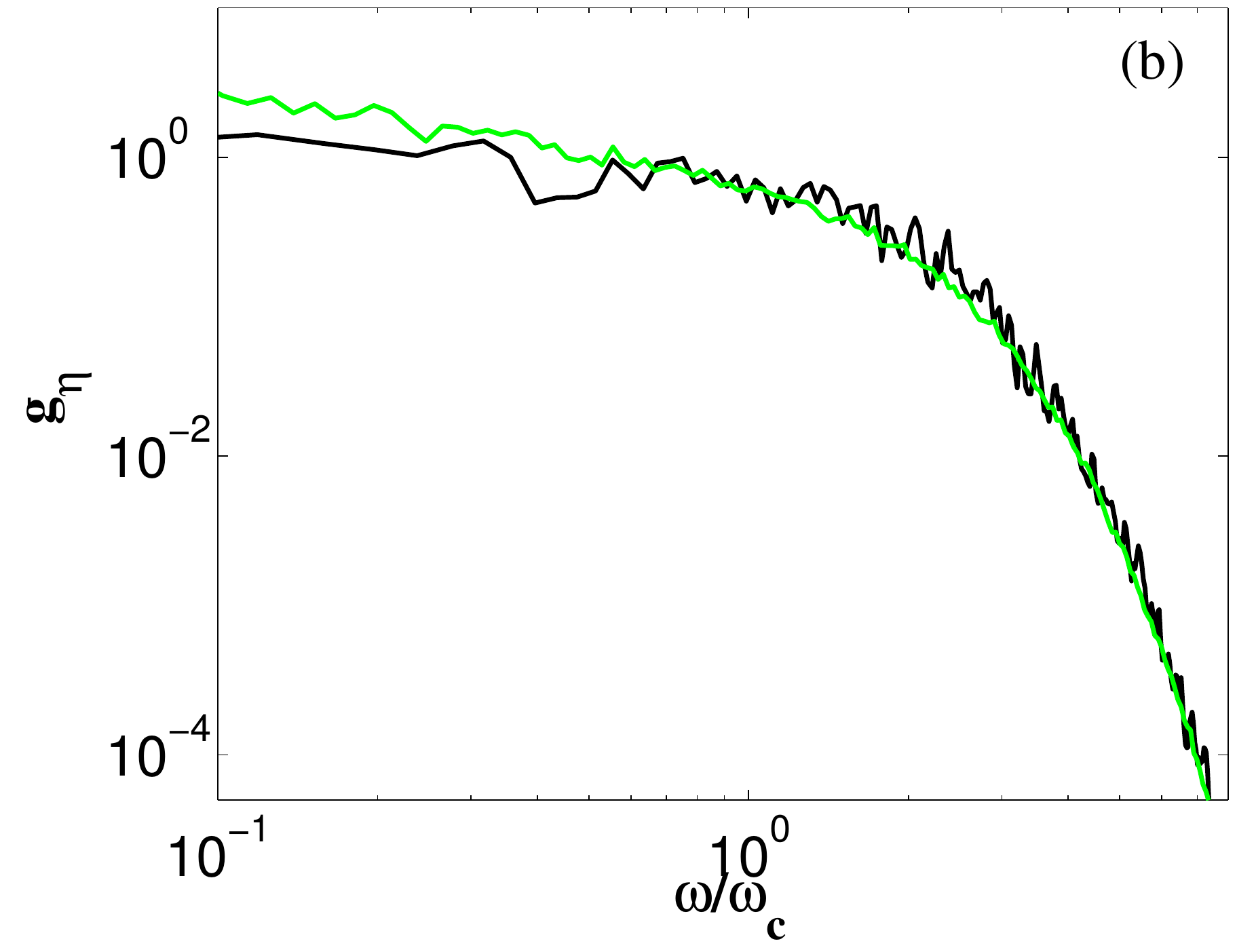}\end{center}
\caption{\label{Fig2} Self-similar function $g_\eta$ in case of non-stationary forced turbulence. (a) Blue points: numerical simulation of Eq.~\eqref{Eq.3} with $\varepsilon_I=1$. Red dashed line: solution of Eq.~\eqref{Eq.10}. (b) Direct simulations of the F\"{o}ppl-von K\'arm\'an equations. Black line: finite-difference and energy-conserving scheme~\cite{humbert_ref21}. Green line: pseudo-spectral method detailed in~\cite{humbert_ref11}.}
\end{figure}

Fig.~\ref{Fig2}(a) compares the self-similar solution deduced from the phenomenological model (and already displayed in Fig.~\ref{Fig1}(c)) with the self-similar solution provided by Eq.~\eqref{Eq.10}. A perfect agreement is observed, exhibiting in particular a cut-off above which the solution vanishes. As shown in \cite{humbert_ref21}, the self-similar solution can be obtained directly from the kinetic equation. However in this case, the general shape of the function is not provided by the theory. Thanks to Eq.~\eqref{Eq.10}, the phenomenological model is able to predict the shape of the self-similar function.  

Let us now compare this solution with direct numerical simulations.
Fig.~\ref{Fig2}(b) shows the obtained results, rescaled according to the self-similar relationship proposed in Eq.~\eqref{Eq.10}. Two different numerical schemes have been used for a better comparison. On the one hand, a finite-difference and energy-conserving scheme  simulates a perfect rectangular plate with simply-supported out-of-plane boundary conditions and in-plane movable edges \cite{humbert_ref21}. The plate has a surface of 0.4$\times$ 0.6 m$^2$, the thickness is 1 mm, and the material parameters are that of a metal, see \cite{humbert_ref21} for more details. The other solution is obtained thanks to the pseudo-spectral method used in previous works~\cite{humbert_ref11,humbert_ref15}, where such a spectral approach leads to periodic boundary conditions. The simulated plate has also the material properties of a metal and corresponds to a square of $0.4\times 0.4$ $m^2$ and its thickness is $1$ mm~\cite{humbert_ref11}. In both numerics, the plate is continuously excited at large scale, corresponding roughly to a constant injection of energy with time. For the finite-difference simulation, this is realized with a pointwise forcing, the frequency of which is selected in the vicinity of the fourth eigenfrequency. 
For the pseudo-spectral code, this is realized in the Fourier space directly through a random noise acting at small wave numbers only. With the two numerical schemes, a clear self-similar behaviour has been observed. Hence we are in position to compare the master curves of the self-similar process for the phenomenological model with those found in the numerical simulations. For the detailed presentation of the self-similar process found in direct numerical simulations, the interested reader is referred to~\cite{humbert_ref21}.

Fig.~\ref{Fig2}(b) shows that the two different numerical methods exhibit similar rescaled spectra. 
Comparing to Fig.~\ref{Fig2}(a), one can observe two discrepancies between the two solutions:
\begin{itemize}
\item In the direct numerical simulations, the slope in the turbulent cascading regime is a bit steeper. This can be assigned to the presence of the forcing term in the very-low frequency part of the spectrum, which creates a small prominence that has already been observed and commented, see {\em e.g.} \cite{humbert_ref21,humbert_ref26}.
\item Near the cut-off, it appears that numerical spectra of the full dynamics decrease continuously and smoothly, whereas theoretical spectra display a steep cut-off because of the non-existence of the mathematical solution. This observation is similar to what has been obtained for the KZ stationary spectrum.
\end{itemize}
Despite these two differences, the general shape of the self-similar solutions in the case of non-stationary forced turbulence shows a very good agreement, validating the results provided by the phenomenological model.

\subsection{Free Turbulence}

The case of free turbulence, {\em i.e.} the evolution of the cascade without external forcing, for a given amount of energy as initial condition, is now considered. As shown in~\cite{humbert_ref21} from the kinetic equation and confirmed  by direct numerical simulation, the cascade front must evolve to high frequencies as $t^{1/3}$. The ability of the phenomenological model to retrieve this dynamics is now investigated.

\subsubsection{Self-similar analysis}

Considering free turbulence leads to withdraw forcing and damping terms. The system being conservative, the amount of initial energy $K$ is conserved, so that Eq.~\eqref{Eq.8} is replaced by:
\begin{equation}
\int_0^{+\infty}{E_\omega d\omega} = K.
\end{equation}
The second relationship that links the unknowns $\alpha$ and $\beta$ now turns to be $\alpha=-\beta$, leading to $\alpha = -1/3$ and $\beta=1/3$. The self-similar solution for the energy spectrum $E_\omega$ reads in this case
\begin{equation}
E_\omega = t^{-1/3}h(\frac{\omega}{t^{1/3}}).
\label{Eq.13}
\end{equation}

\begin{figure}
\begin{multicols}{2}
\includegraphics[width=6cm,height = 5.5cm]{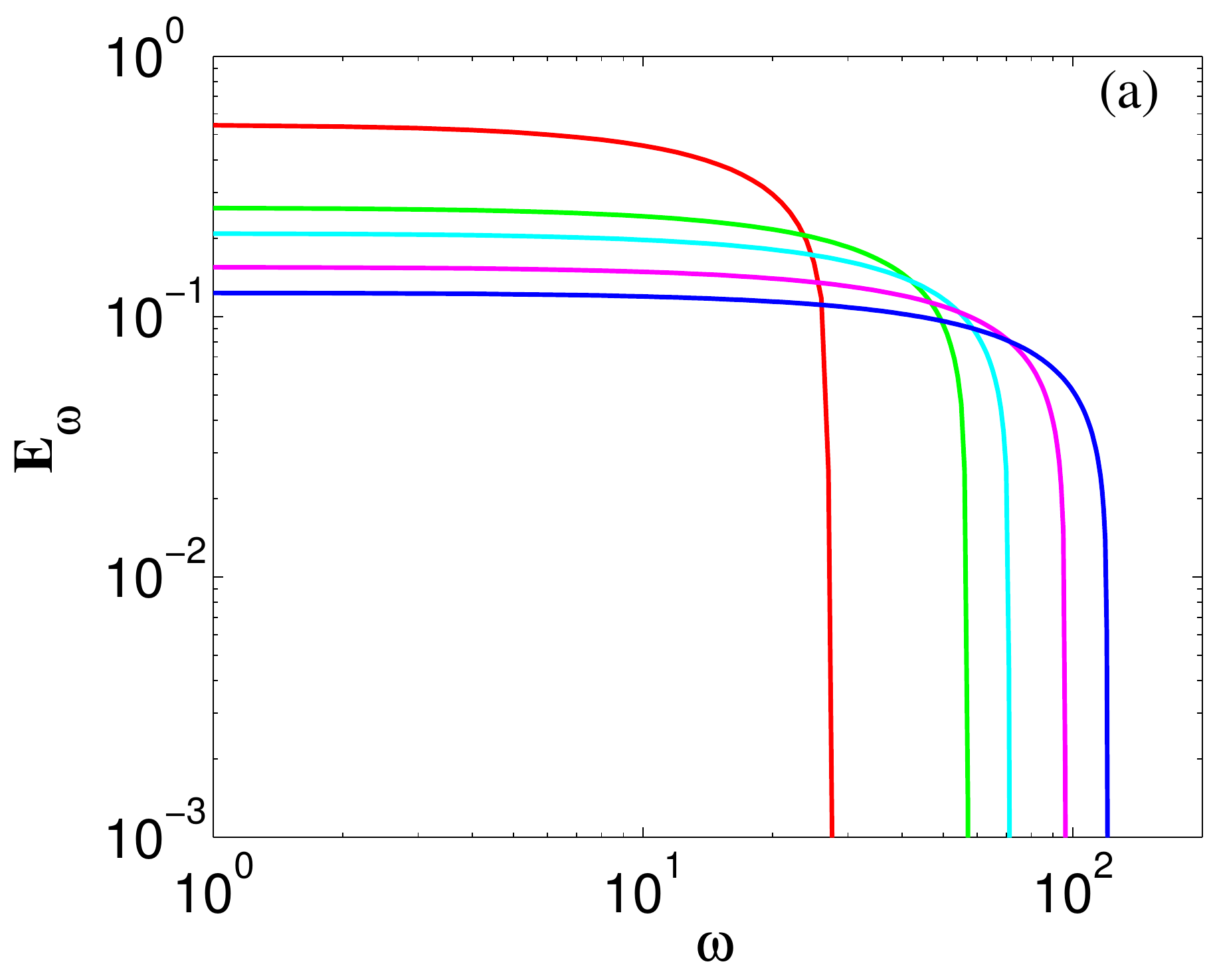}
\includegraphics[width=4.5cm, height=3cm]{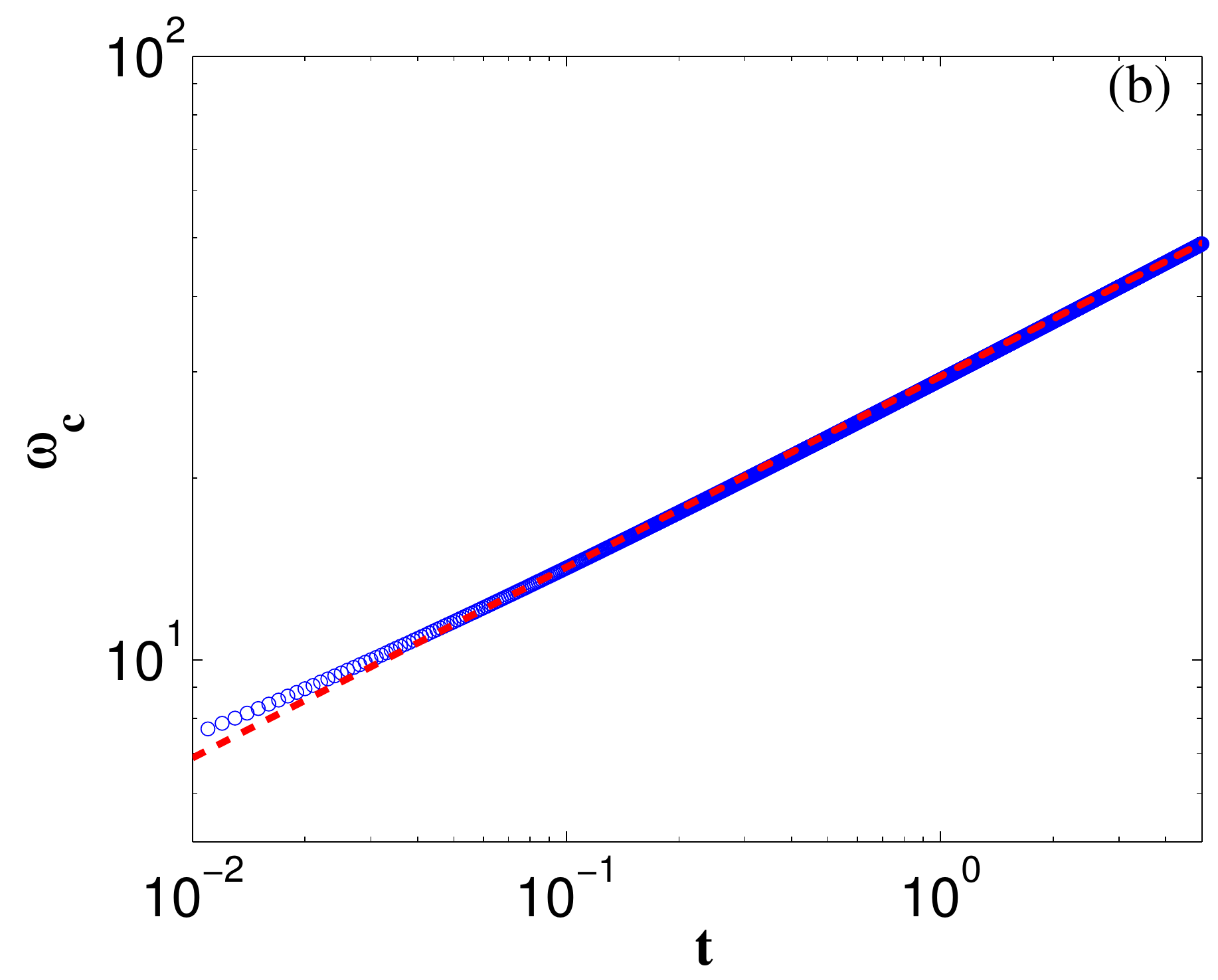}\\
\includegraphics[width=4.5cm, height=3cm]{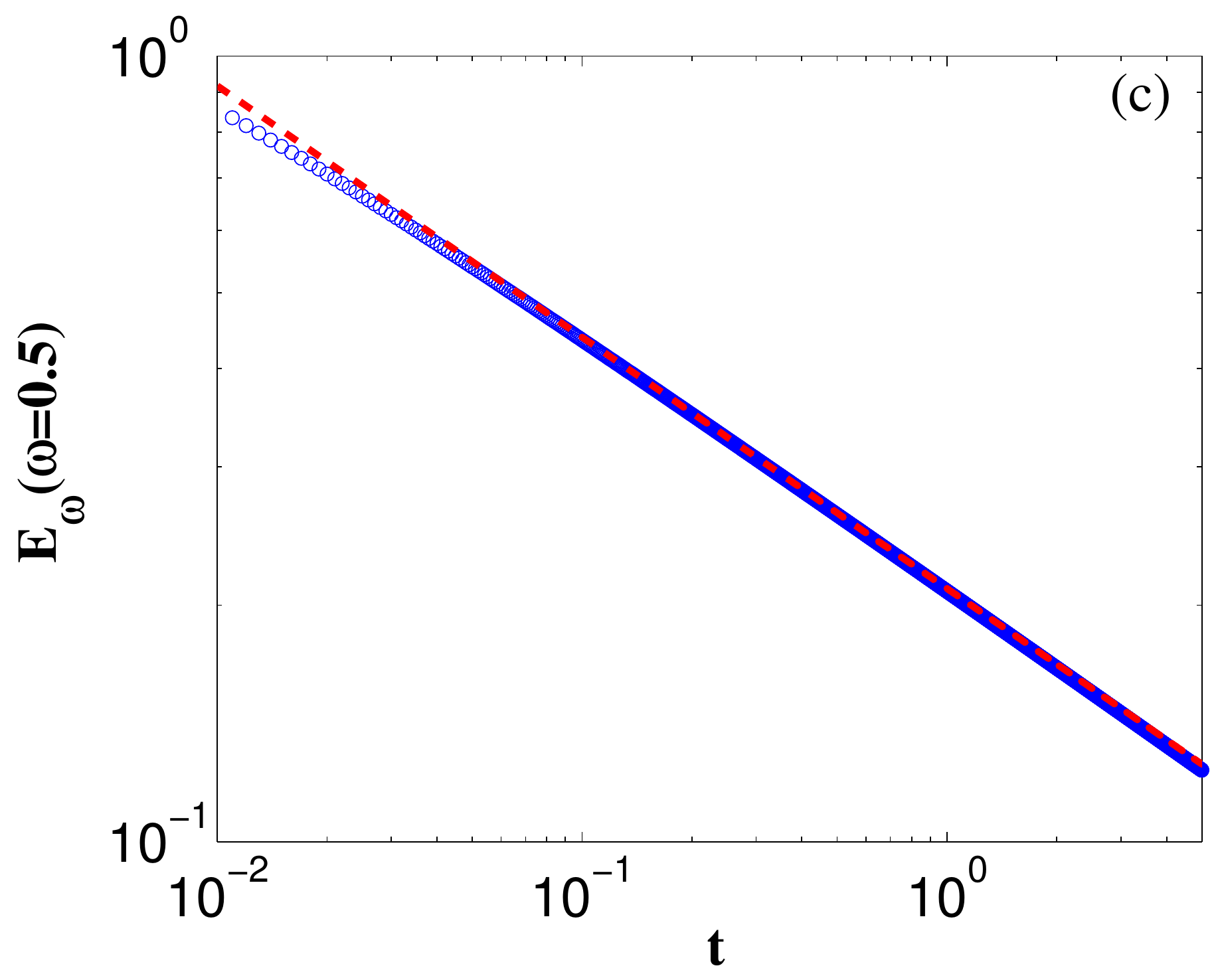}
\end{multicols}
\caption{\label{Fig3} (a) Energy spectrum $E_\omega$ as a function of the frequency $\omega$ at (from top to bottom) $t = 0,1,2,3,4$ [nondim]. (b) Characteristic frequency $\omega_c$ as a function of time. Red dashed line: $\omega_c \propto t^{1/3}$. (c) $E_\omega(\omega=0.5)$ as a function of time. Red dashed line: $E_\omega(\omega=0.5)\propto t^{-1/3}$.}
\begin{center}\includegraphics[width=0.43\textwidth]{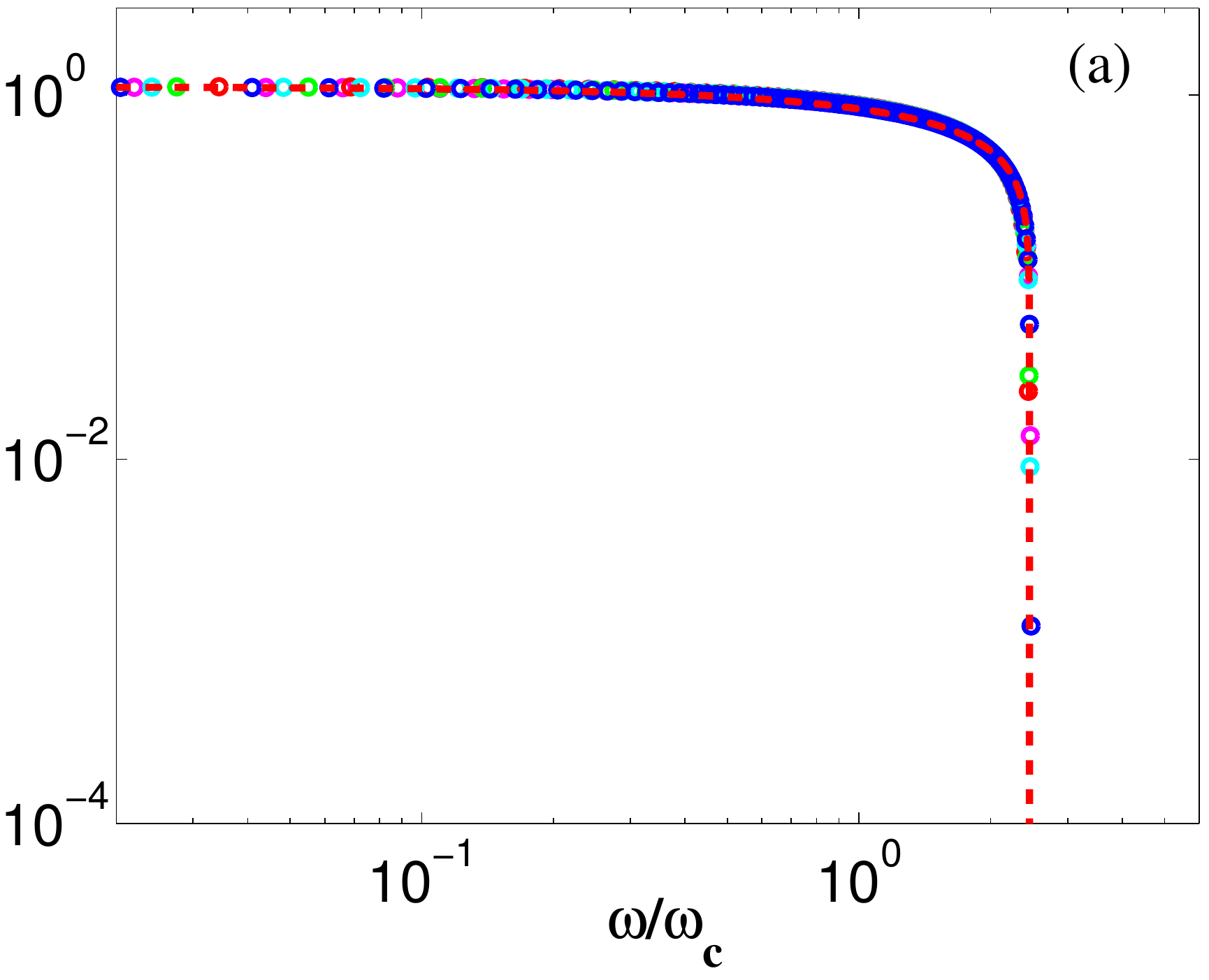}
\hspace{0.5cm}
\includegraphics[width=0.45\textwidth]{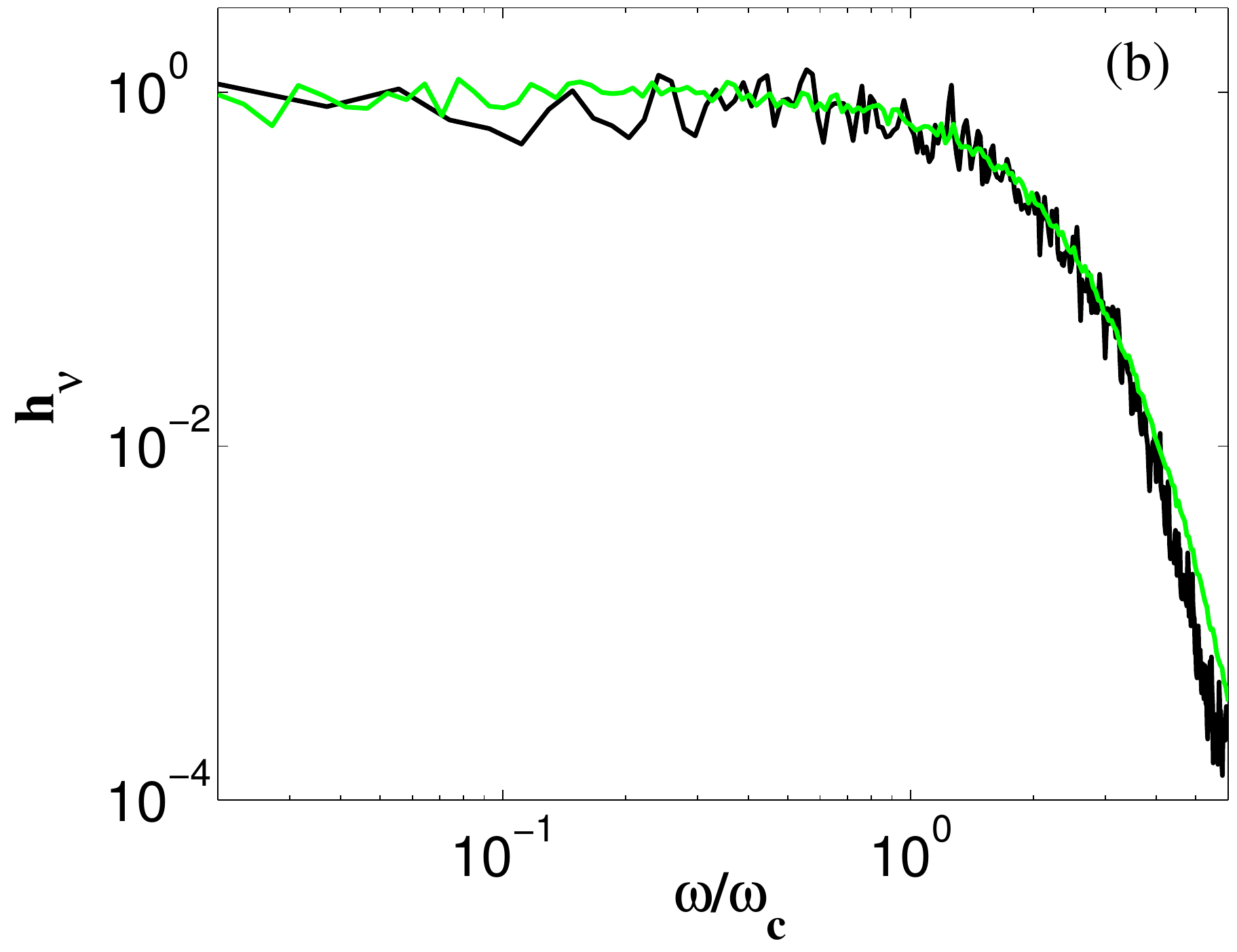}\end{center}
\caption{\label{Fig4} Self-similar function in case of free turbulence. (a) Points: spectra of Fig.~\ref{Fig3} rescaled by the self-similar law given by Eq.~\eqref{Eq.13}. Red dashed line: solution of Eq.~\eqref{Eq.14}. (b) Results of the direct simulations of the F\"{o}ppl-von K\'arm\'an equations. Black line: finite-difference and energy-conserving scheme~\cite{humbert_ref21}. Green line: pseudo-spectral method detailed in~\cite{humbert_ref11}.}
\end{figure}

In order to simulate numerically the framework of free turbulence, the dissipation introduced earlier at high frequency, is now removed. An energy flux $\varepsilon_I$ is imposed for a few time steps and then cancelled, thus fixing the origin of time. Then, the simulation is run by imposing a vanishing energy flux at $\omega = 0$, ensuring free turbulence. Fig.~\ref{Fig3}(a) shows the evolution for an initial amount of energy K (corresponding to the spectrum in red) as a function of time. Fig.~\ref{Fig3}(b)(c) describe the evolution of the characteristic frequency $\omega_c$ as well as the evolution of the amplitude of the energy spectrum at the centre of the first mesh element ($\omega=0.5$). Two behaviours respectively proportionals to $t^{1/3}$ and to $t^{-1/3}$ are displayed. These two observations are in agreement with the self-similar solution given by Eq.~\eqref{Eq.13}. 

In the same manner as for the forced case, the solution given by Eq.~\eqref{Eq.13} can be inserted into Eq.~\eqref{Eq.3} in order to obtain the evolution equation of the self-similar function $h_\nu = h(\omega/t^{1/3})$. The analogue of Eq.~\eqref{Eq.10} for the free turbulence case then reads 
\begin{equation}
-\frac{1}{3}\nu h_\nu' = (\nu h_\nu^2h'_\nu)',
\label{Eq.14}
\end{equation}
where $'$ stands here for the derivative with respect to the self-similar variable $\nu=\omega/t^{1/3}$. The numerical method used in order to solve Eq.~\eqref{Eq.10} is now applied to Eq.~\eqref{Eq.14}. Fig.~\ref{Fig4}(a) displays the self-similar function built from the spectra calculated by the phenomenological model at multiple times and scaled as prescribed by Eq.~\eqref{Eq.13}. For comparison, the solution of the self-similar equation Eq.~\eqref{Eq.14} is also represented. A good agreement is observed, confirming the self-similar evolution of the spectrum. Fig.~\ref{Fig4}(b) displays the numerical results from the direct numerical simulations of the F\"{o}ppl-von K\'arm\'an equations. Once again, the two numerical schemes leads to functions that are very close from each other. A much better agreement is observed between the solutions from direct simulations and the one from the phenomenological model, in particular the slope in the cascade regime are really the same. This confirms once again the effect of the forcing which creates a small bump in the very low-frequency part of the spectrum and alters the direct comparison between the different solutions. Here in the free turbulence case, a perfect agreement is observed, the only difference being the behaviour near the cut-off frequency where the decrease of the spectrum is much slower for the direct numerical simulations, as already commented.\\

Two situations belonging to the theoretical conservative framework of wave turbulence in thin vibrating plates have been investigated through numerical simulations of the phenomenological equation. Self-similar behaviours pertaining to the phenomenology of the F\"{o}ppl-von K\'arm\'an equations have been successfully recovered. Note also that the behaviours $E _\omega= g(\frac{\omega}{t})$ for forced turbulence and $E_\omega = t^{-1/3}h(\frac{\omega}{t^{1/3}})$ for free turbulence can be derived by an analysis of the kinetic equation, as shown in \cite{humbert_ref21}. However, in this case the self-similar functions $g$ and $h$ are left unknown. Thanks to the phenomenological model, two different ordinary differential equations have been deduced, the solutions of which are functions $g$ and $h$. Hence, the model gives further informations which have been found to be relevant by comparisons with the direct numerical simulations. All these results show the ability of our simple equation to recover complex features of the physics of the problem.

\section{Non conservative case : the effect of damping}

\subsection{Model equation}

Physical dissipation can be introduced in the phenomenological model by adding a linear dissipation term to Eq.~\eqref{Eq.3}: 

\begin{equation}
\partial_t E_\omega = \partial_\omega(\omega E_\omega^2\partial_\omega E_\omega) - \hat\gamma E_\omega,
\label{Eq.15}
\end{equation}
where $\hat\gamma$ can be chosen as a function of $\omega$ for the sake of generality. In thin plates, the damping depends strongly on parameters such as the size of the plate, its thickness, the boundary conditions. Regarding these values and the frequency range of interest, either thermoelastic, viscoelastic, acoustical radiation, or losses through the boundary conditions, can dominate~\cite{Norris_06, Chaigne_01, Arcas_09, Caracciolo_95}. In the framework of our experimental set-up, the importance of most of these contributions has been estimated and related to theoretical predictions in~\cite{Humbert_14}.

As a starting point, let us consider the damping laws obtained from experiments. As observed in~\cite{humbert_ref15} where experimental methods have been used in order to increase the amount of damping in the plate, the damping laws for four different configurations were found to follow the power-law $\hat\gamma = \xi \omega^{0.6}$, with relative values of $\xi$ (with respect to the smallest one) ranging from 1 to 5. This damping law with varying $\xi$ is first used for investigating the solutions of Eq.~\eqref{Eq.15}. Appendix B gives the full correspondence between experimentally measured values of $\xi$ and their respective dimensionless counterparts used in the numerical simulations of Eq.~\eqref{Eq.15}. The same finite volume method is used  as in the previous sections, and the flux of energy $\varepsilon_I$ is fixed at $\omega=0$. After a certain number of time steps (depending on the selected damping coefficient $\xi$), a stationary regime is reached.

\begin{figure}
\begin{center}\includegraphics[width=0.8\textwidth]{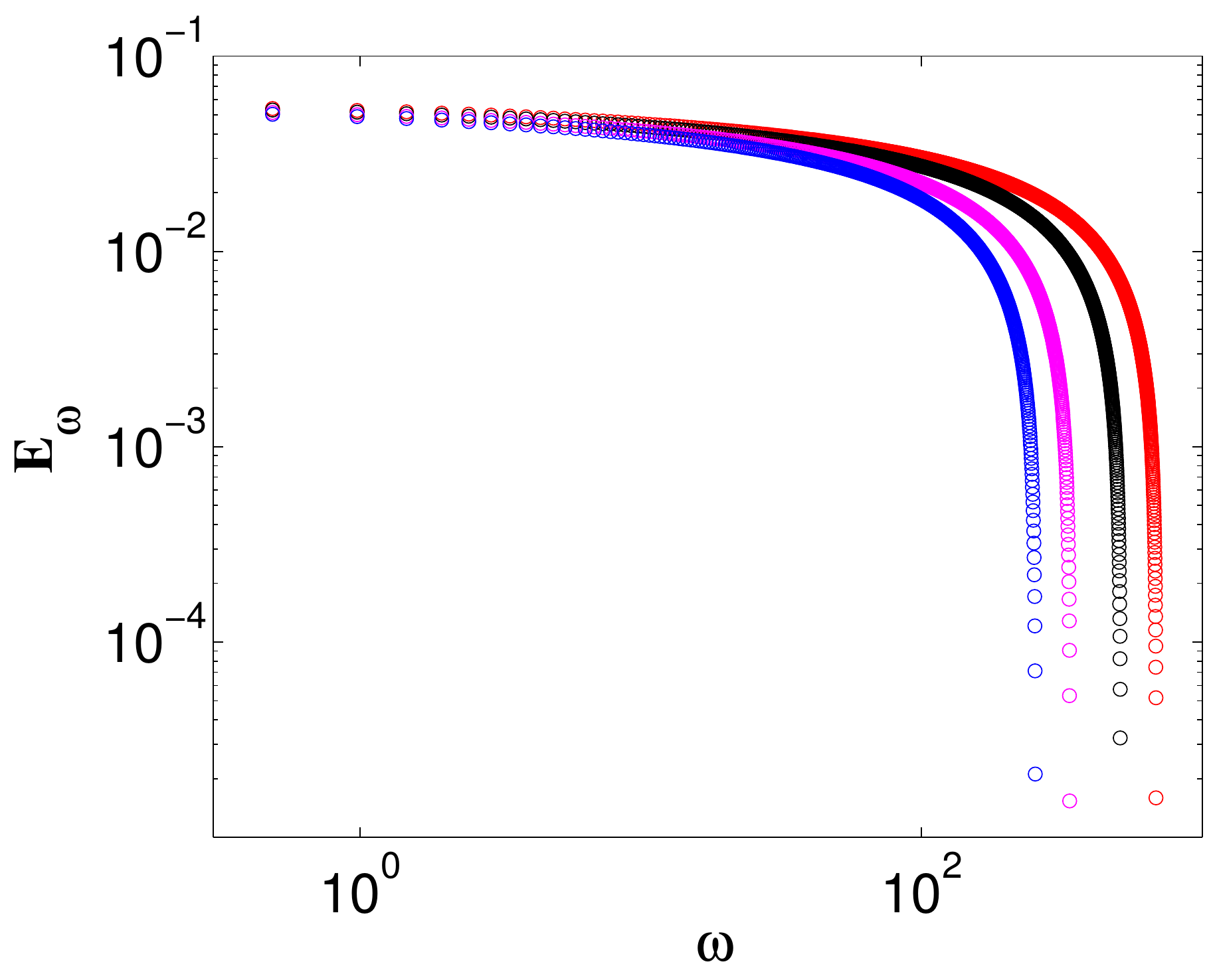}\end{center}
\caption{\label{Fig5} Stationary energy spectrum $E_\omega$ in the damped case, as a function of the frequency $\omega$ and for $\varepsilon_I = 1\times10^{-5}$. Red: $\xi = 1.908\times 10^{-5}$. Black: $\xi = 3.0528\times 10^{-5}$. Magenta: $\xi = 5.9359\times 10^{-5}$. Blue: $\xi = 9.3279\times 10^{-5}$}.
\begin{center}\includegraphics[width=0.8\textwidth]{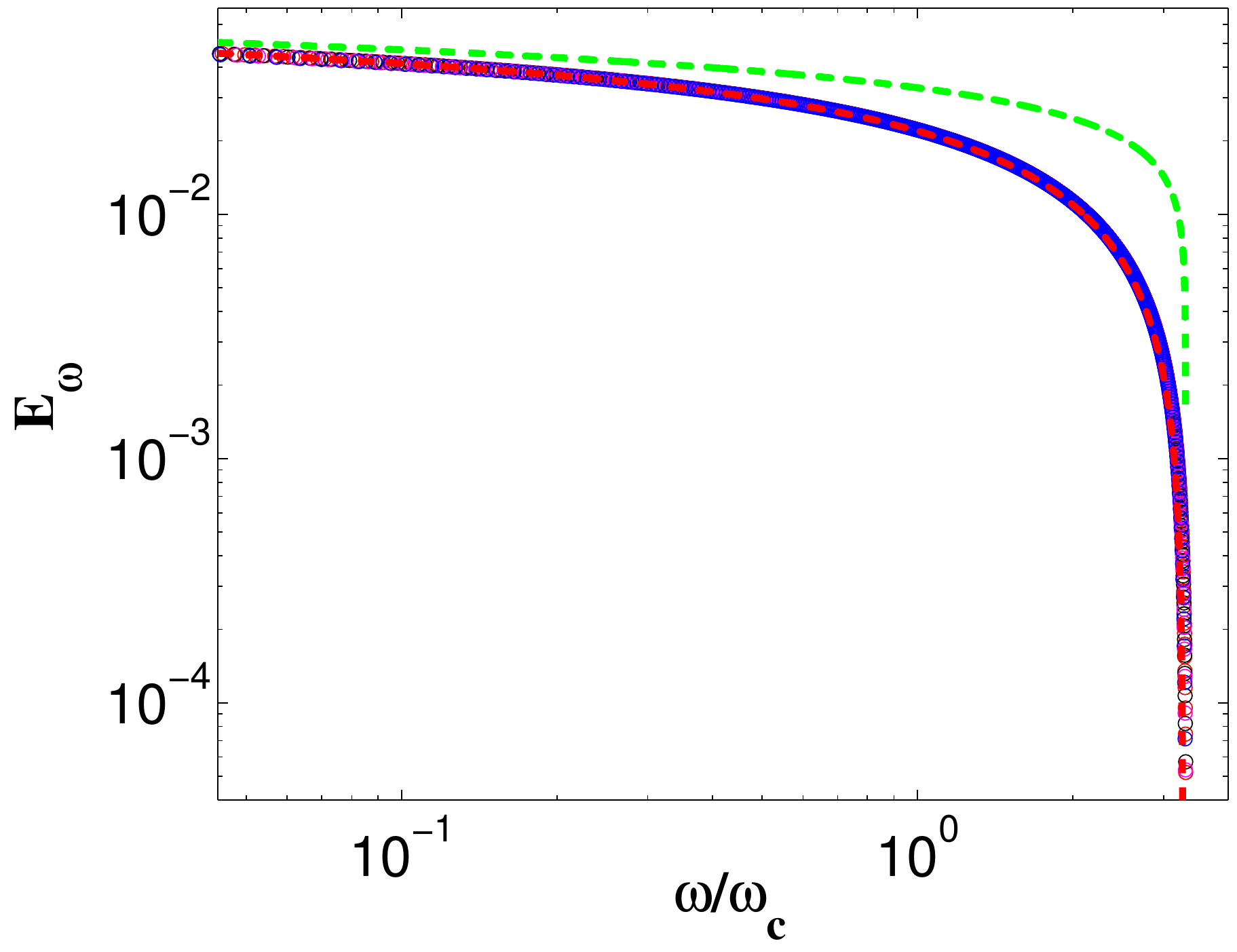}\end{center}
\caption{\label{Fig6}Energy spectra displayed in Fig.~\ref{Fig5} as functions of the rescaled frequency $\omega/\omega_c$. Green dashed line: Kolmogorov-Zakharov spectrum $E^{KZ}_\omega = (3\varepsilon_I)^{1/3}\log(\frac{\omega^{\star}}{\omega})^{1/3}$ for $\varepsilon_I = 1\times10^{-5}$. Red dashed line: solution of Eq.~\eqref{Eq.22}. }
\end{figure}

Fig.~\ref{Fig5} exhibits the stationary energy spectrum obtained for each of the four damping cases retrieved from~\cite{humbert_ref15}, with an amount of damping coefficient $\xi$ multiplied by 5 between the smallest and largest ones. The energy flux at $\omega=0$ is the same for each situation. For very low frequencies (say $0<\omega<5$), all spectra shows roughly the same behaviour. For larger frequencies, the more damped the system, the steeper the spectrum and the smaller its characteristic frequency are. Moreover, it appears that the dissipation affects the energy transfers between scales, so that summing up the stationary spectra to power laws is not possible anymore. 

Fig.~\ref{Fig6} shows the previous spectra as functions of the rescaled frequency $\omega/\omega_c$. The rescaling of the frequency axis makes all spectra collapse into a single curve, which appears to be steeper than the Kolmogorov-Zakharov spectrum (displayed by a green dashed line in Fig.~\ref{Fig6}). This result, obtained with the phenomenological model, is similar to the conclusions already reported in~\cite{humbert_ref15} from experiments only: damping plays an important role in the discrepancies between theoretical and experimental spectra. However, this unique master curve has never been observed before and tends to provide a simple explanation on the behaviour of the cascade in presence of damping. Indeed, it shows that the effect of damping on the turbulent cascade can  be mainly attributed to the balance between the conservative term $\partial_\omega(\omega E^2_\omega\partial_\omega E_\omega)$ and the dissipative term $\hat\gamma_\omega E_\omega$, since only these terms are present in the phenomenological model, and allows one to retrieve the experimental observations. There is obviously no inertial range so that the stationary solution depends on the shape of the dissipation function and differs from the Kolmogorov-Zakharov spectrum. 

Finally, the collapse suggests a self-similar behaviour of the spectrum as a function of the injected flux $\varepsilon_I$ and the damping coefficient $\xi$. In order to derive the equation corresponding to this self-similar solution, the energy spectrum $E_\omega$ is thus written under the form
\begin{equation}
E_\omega = \varepsilon_I^\mu{\xi}^xf_\eta\left(\frac{\omega}{\omega_c}\right)\;\;\mbox{with}\;\;\omega_c = \varepsilon_I^y{\xi}^z,
\label{Eq.16}
\end{equation}
where $f_\eta$ is an unknown function of the self-similar variable $\eta = \omega/\omega_c$ and $\mu, x, y, z$ are constants to be determined. Recalling that the injected flux $\varepsilon_I$ corresponds in the phenomenological model to
\begin{equation}
\varepsilon_I=\lim_{\omega\rightarrow0} (-\omega E^2_\omega\partial_\omega E_\omega),
\label{Eq.17}
\end{equation}
one obtains, after inserting Eq.~\eqref{Eq.16} into Eq.~\eqref{Eq.17}, the following relationship:
\begin{equation}
\varepsilon_I=-\varepsilon_I^{3\mu}\xi^{3x}\lim_{\eta\rightarrow0} (\eta f_\eta^2\partial_\eta f_\eta),
\end{equation}
so that $\mu=1/3$ and $x=0$. The energy spectrum must thus write:
\begin{equation}
E_\omega = \varepsilon_I^{1/3}f_\eta\left(\frac{\omega}{\varepsilon_I^y{\xi}^z}\right).
\end{equation}
In addition, inserting Eq.~\eqref{Eq.16} in the phenomenological equation~\eqref{Eq.15} with a damping of the form of an unknown power law  $\hat{\gamma}= \xi\omega^{\lambda}=\xi\eta^\lambda\omega_c^\lambda$ yields:
\begin{equation}
\partial_tE_\omega=0=\varepsilon_I^{1-y}\xi^{-z}\partial_\eta(\eta f_\eta^2\partial_\eta f_\eta)-\varepsilon_I^{\lambda y+1/3}\xi^{\lambda z+1}\eta f_\eta.
\end{equation}
Thus, the unknowns $y$ and $z$ must fulfil the following relationships that depends on the frequency dependence of the damping:
\begin{equation}
z=-\frac{1}{1+\lambda},\qquad\qquad y=\frac{2}{3(1+\lambda)}.
\end{equation} 
All the unknowns of Eq.~\eqref{Eq.16} have been determined, leading to an equation for the function $f_\eta$,
\begin{equation}
\partial_\eta(\eta f_\eta^2\partial_\eta f_\eta)-f_\eta\eta^\lambda  = 0,
\label{Eq.22}
\end{equation}
and to an expression for the characteristic frequency as a function of the damping and the injected flux:
\begin{equation}
\omega_c=\varepsilon_I^\frac{2}{3(1+\lambda)}\xi^{-\frac{1}{1+\lambda}}.
\label{Eq.23}
\end{equation}
Eq.~\eqref{Eq.22} has no analytical solution but can be solved numerically following the same procedure as for Eq.~\eqref{Eq.10} and \eqref{Eq.14}. The result is plotted in red in Fig.~\ref{Fig6}, displaying a perfect agreement with the universal solution obtained by rescaling all the spectra.

\begin{figure}
\begin{center}\includegraphics[width=8.6cm]{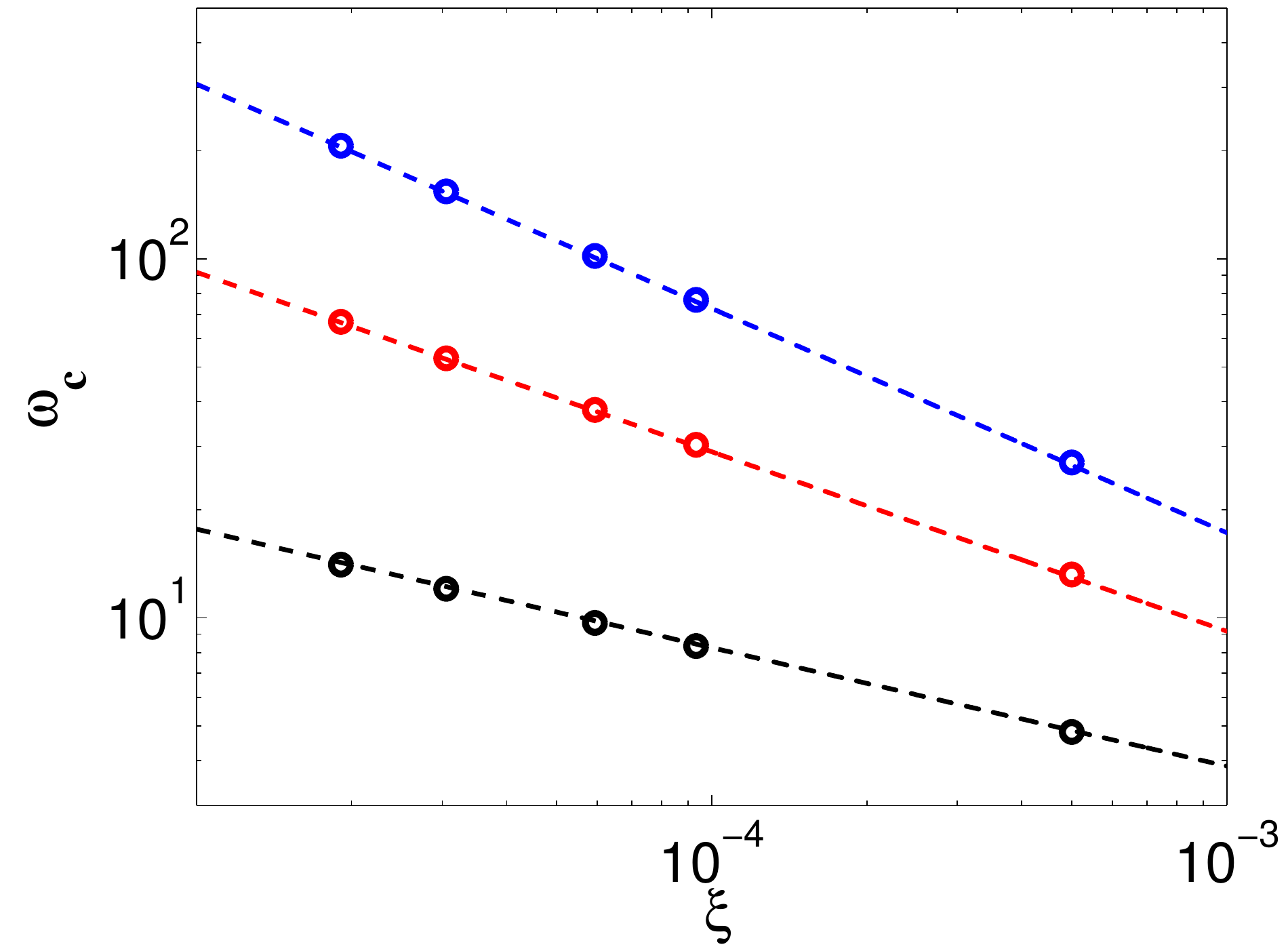}\end{center}
\caption{\label{Fig7}Characteristic frequency $\omega_c$ as a function of the damping coefficient $\xi$, $\varepsilon_I = 1\times10^{-5}$. Black: $\lambda = 2$. Red: $\lambda = 1$. Blue: $\lambda = 0.6$. Dashed lines: evolution laws predicted by Eq.~\eqref{Eq.23}.}
\end{figure}

To conclude this part, the validity of Eq.~\eqref{Eq.23}, which expresses the behaviour of the characteristic frequency, is questioned. As already observed in Fig.~\ref{Fig5} for $\lambda = 0.6$, increasing the damping coefficient $\xi$ decreases the characteristic frequency $\omega_c$. In this case, the theoretical prediction provided by Eq.~\eqref{Eq.23} reads
\begin{equation}
\omega_c = \varepsilon_I^{5/12}\xi^{-5/8}.
\end{equation}
Fig.~\ref{Fig7} compares this prediction with the characteristic frequencies obtained by solving Eq.~\eqref{Eq.15} for $\hat\gamma=\xi\omega^\lambda$ and $\lambda = 0.6$. The same study for $\lambda = 1$ and $\lambda =2$ is also displayed. A perfect agreement is found, showing that the evolution of the characteristic frequency can be fully explained thanks to the self-similar behaviour of the energy spectrum with damping and injected flux.

\subsection{Discussion} 

The results of the previous section, obtained with the phenomenological model, have shown the existence of a unique master curve on which all spectra collapse when rescaling the frequency with respect to the characteristic frequency. This feature has not been noticed before in the experimental results reported in~\cite{humbert_ref15}, where four different configurations of damping for the same plate have been measured. It is potentially a very important result since it suggests that the change in the cascade slope observed when the damping varies (following the stronger the dissipation, the steeper the energy spectra are) is simply a consequence of the master curve which does not exhibit a single slope. Depending on the dissipation a different region of the master curve is dominating, exhibiting different "apparent" slope. It is thus crucial to investigate whether this feature is also present in experiments and in numerical simulations of the plate equations.

\begin{figure}
\begin{center}
\includegraphics[width=6cm, height=5cm]{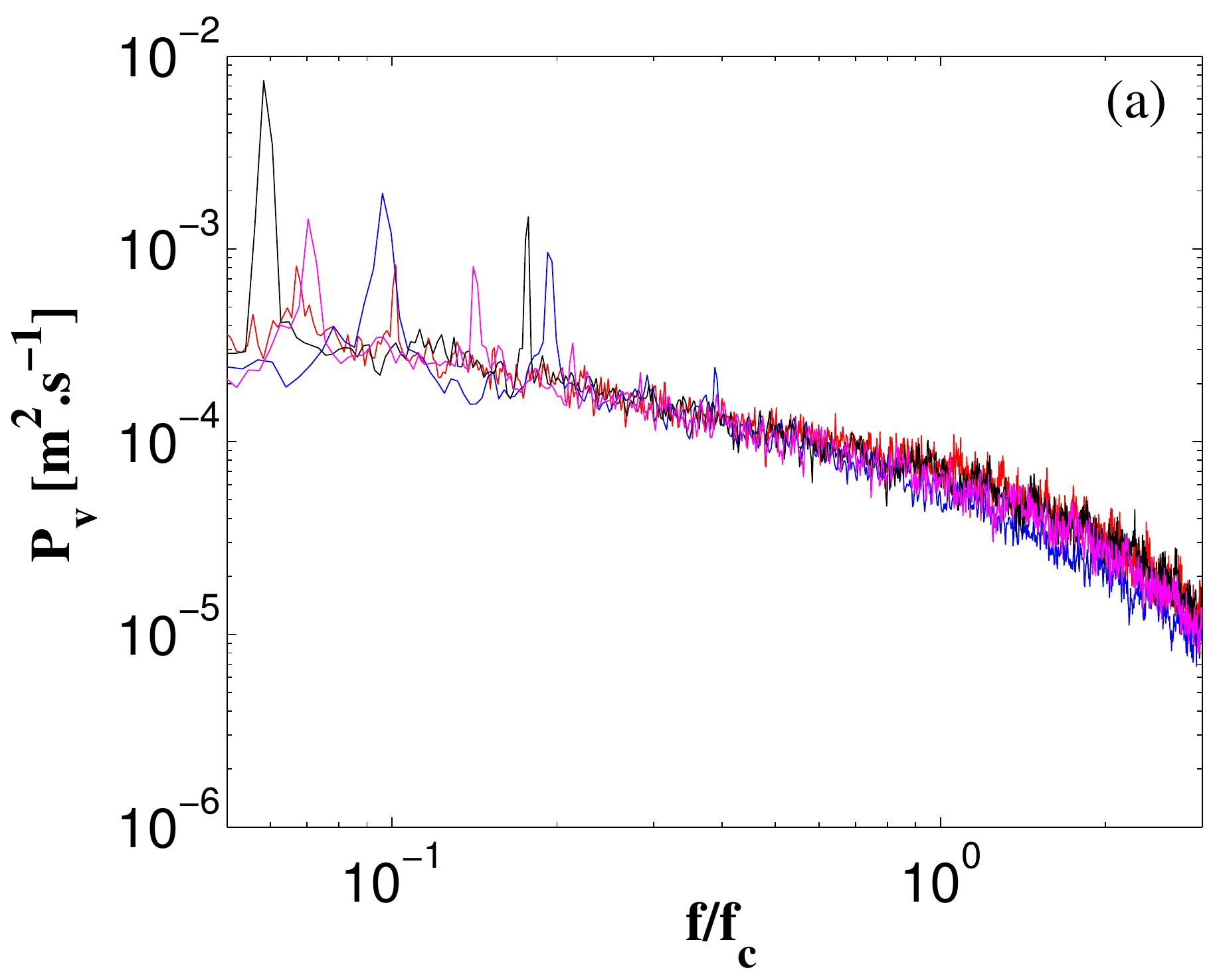}
\includegraphics[width=6cm, height=5cm]{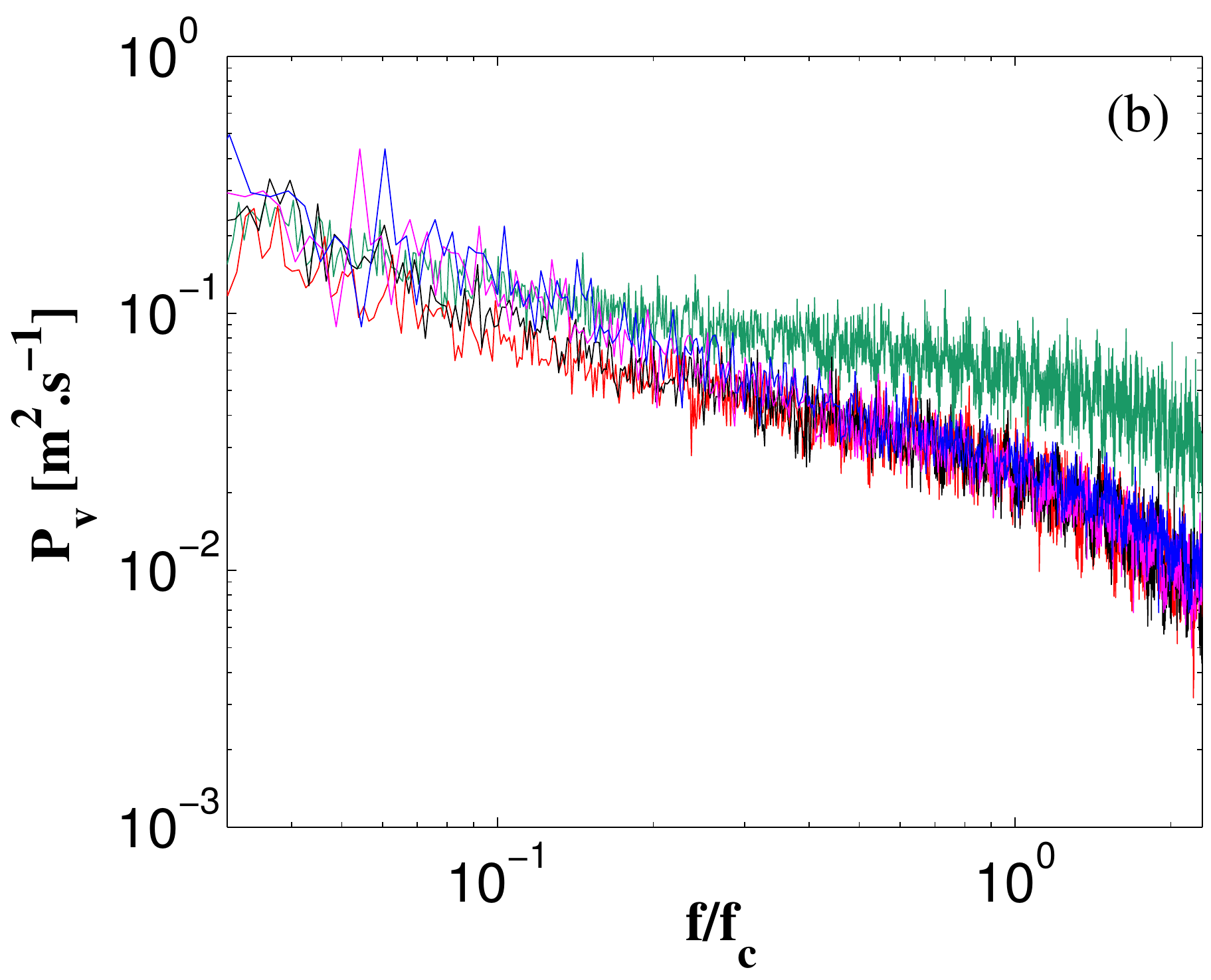}
\end{center}
\caption{\label{Fig8} Power spectral density of the transverse velocity $P_v$ as a function of the rescaled frequency $f/f_c$. Red: $\xi=0.045$. Black: $\xi=0.072$. Magenta: $\xi=0.14$. Blue: $\xi=0.22$. (a) Experiments. Red: $\varepsilon_I=0.56\times10^{-3}$~$\mbox{m}^3.\mbox{s}^{-3}$. Black: $\varepsilon_I=0.54\times10^{-3}$~$\mbox{m}^3.\mbox{s}^{-3}$. Magenta: $\varepsilon_I=0.52\times10^{-3}$~$\mbox{m}^3.\mbox{s}^{-3}$. Blue: $\varepsilon_I=0.48\times10^{-3}$~$\mbox{m}^3.\mbox{s}^{-3}$. (b) Numerical simulations. Green: $\xi=0$, $\varepsilon_I=0.057\times10^{-3}$~$\mbox{m}^3.\mbox{s}^{-3}$. Other cases: $\varepsilon_I=0.024\times10^{-3}$~$\mbox{m}^3.\mbox{s}^{-3}$.}
\end{figure}

Fig.~\ref{Fig8} displays precisely the experimental and numerical (pseudo-spectral method) power spectral densities $P_v$ from~\cite{humbert_ref15} as functions of the rescaled frequency $f/f_c$ for different damping coefficients $\xi$. As for the phenomenological model, the proposed rescaling causes all curves to collapse into a unique master curve. In Fig.~\ref{Fig8}(b),  the spectra from the damped case are compared to the KZ spectrum obtained numerically when the dissipation is only located at high frequency, showing that the spectra are clearly steeper than the usual KZ spectrum.
Moreover, both the experimental and the numerical cases exhibit similar profiles, but are very
different from the master curve of the phenomenological model, in the same vein than the other situations studied above. Nevertheless, these figures confirm here that the observations brought by the phenomenological model describe a true feature of the physical system.

\begin{figure}[h!]
\begin{center}\includegraphics[width=7.6cm]{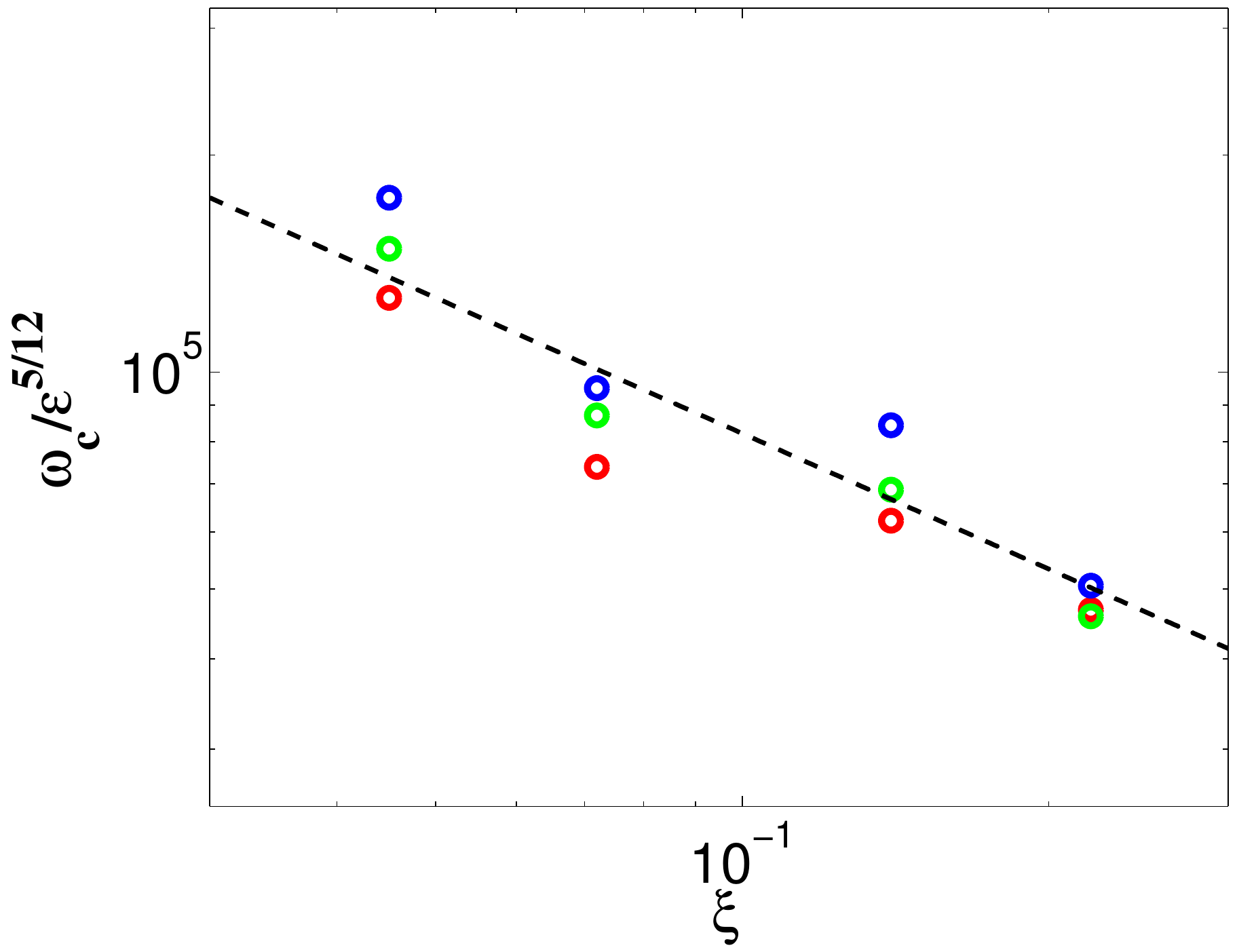}\end{center}
\caption{\label{Fig9}Ratio $\omega_c/\varepsilon_I^{5/12}$ as a function of the damping coefficient $\xi$. $\lambda = 0.6$. Red: $\varepsilon_I=0.52\times10^{-3}$~$\mbox{m}^3.\mbox{s}^{-3}$. Green: $\varepsilon_I=0.16\times10^{-3}$~$\mbox{m}^3.\mbox{s}^{-3}$. Blue: $\varepsilon_I=0.58\times10^{-4}$~$\mbox{m}^3.\mbox{s}^{-3}$. Line: evolution law predicted by Eq.~\eqref{Eq.23}: $\omega_c/\varepsilon_I^{5/12} \propto {\xi}^{-5/8}$.}
\end{figure}

Finally, the relation Eq.~\eqref{Eq.23} between the characteristic frequency, the damping and the injected power can also be questioned using the experimental results. Fig.~\ref{Fig9} displays, for three injected powers, the evolution of the ratio $\omega_c/\varepsilon_I^{5/12}$ as a function of the damping parameter $\xi$. The predicted dependence of $\omega_c$ with $\xi$ is also drawn for comparison: $\omega_c/\varepsilon_I^{5/12}\propto{\xi}^{-5/8}$. The accordance is good, confirming that the results of the model are in agreement with the behaviour of the experiments.

\section{Conclusion}

A phenomenological model describing the time-frequency dependence of the power spectrum for wave turbulence in thin vibrating plates, has been derived. In the framework of non-stationary turbulence, the model equation has shown its ability in predicting the self-similar behaviours for two different cases: free and forced turbulence. These two examples show the ability of our model to capture the most salient features of the dynamics of thin elastic plates. The model equation possesses a number of attracting features for further studies, the prominent one being its simplicity in handling complicating effects such as forcing and dissipation. Besides its ability in recovering the self-similar behaviours already derived from the kinetic equation~\cite{humbert_ref21}, a step further has been obtained with the derivation of two equations, \eqref{Eq.10} and  \eqref{Eq.14}, the solutions of which are the self-similar universal functions for the forced and the free cases, which were not provided by the theory developed from the kinetic equation in~\cite{humbert_ref21}. 

The phenomenological model has then been used in order to further investigate the effect of damping on the spectra of turbulence for thin vibrating plates reminding that, in that case, damping acts at all scales and breaks the transparency window required by the wave turbulence theory. Then, no more power-law behaviour can be observed, and the slope of energy spectra does not represent the most important parameter to investigate~\cite{humbert_ref15}. Thanks to the phenomenological model, a self-similar analysis provides new results and makes appear a relationship between the power spectra, the damping law and the injected power. With the model equation and for a given damping law, all curves collapse into a single one when increasing the damping factor, and the characteristic frequency can be directly studied and predicted from the energy budget of the cascade. All these results shed new light on experimentally observed turbulent spectra with damping. This also confirms that the phenomenological model is a useful  tool for studying complicating effects in wave turbulence of plates.

\appendix
\renewcommand\thesection{Appendix A}

\section{Non-dimensional F\"{o}ppl-von K{\'a}rm{\'a}n equations}
\label{Appendix1}
The dynamics of thin vibrating plates is described by the F\"{o}ppl-von K{\'a}rm{\'a}n equations with two unknowns that are the transverse displacement field $\zeta(x,y,t)$ and the Airy stress function $\chi (x,y,t)$. For a thin plate of thickness $h$, made from a material with Poisson ratio $\nu$, density $\rho$ and Young's modulus $E$, the equations of motion read~\cite{humbert_ref24,humbert_ref27,humbert_ref28}
\begin{eqnarray}
\rho h\frac{\partial^2\zeta}{\partial t^2}&=&-\frac{Eh^3}{12(1-\nu^2)}\Delta^2\zeta+{\mathcal L}(\chi,\zeta),\\
\Delta^2\chi&=&-\frac{Eh}{2}{\mathcal L}(\zeta,\zeta).
\end{eqnarray}
The operator ${\mathcal L}$ is bilinear symmetric, and reads in Cartesian coordinates ${\mathcal L}(f,g)=f_{xx}g_{yy}+f_{yy}g_{xx}-2f_{xy}g_{xy}$. 

The following change of variables is applied to obtain dimensionless variables

\begin{equation} {\bf x}'=\frac{{\bf x}}{l},\qquad \zeta'=\frac{\zeta}{l}, \qquad  t'=\frac{t}{\tau}, \qquad  \chi'=\frac{\chi}{C},
\end{equation}

\noindent where the characteristic length $ l= \frac{h}{\sqrt{3(1-\nu^2)}}$, time $ \tau=l\sqrt{\frac{\rho}{E}}$ and $C= E h l^2$, have been introduced. This leads to the following set of non-dimensional dynamical  equations:

\begin{eqnarray}
\frac{\partial^2\zeta}{\partial t^2}&=&-\frac{1}{4}\Delta^2\zeta+{\mathcal L}(\chi,\zeta),\\
\Delta^2\chi&=&-\frac{1}{2}{\mathcal L}(\zeta,\zeta).
\end{eqnarray}

\renewcommand\thesection{Appendix B}

\section{Correspondence between experimental and phenomenological values of the damping coefficient $\xi$}
\label{Appendix2}
In the F\"{o}ppl-von K{\'a}rm{\'a}n equations, viscous dissipation can be taken into account with the term $\rho h \gamma \frac{\partial \zeta}{\partial t}$, so that the equations of motion writes:

\begin{eqnarray}
\rho h\frac{\partial^2\zeta}{\partial t^2}&=&-\frac{Eh^3}{12(1-\nu^2)}\Delta^2\zeta+{\mathcal L}(\chi,\zeta)- \rho h \gamma \frac{\partial \zeta}{\partial t},\\
\Delta^2\chi&=&-\frac{Eh}{2}{\mathcal L}(\zeta,\zeta),
\end{eqnarray}
where $\gamma$ is the damping factor. The equivalent set of non-dimensional equations becomes:

\begin{eqnarray}
\frac{\partial^2\zeta}{\partial t^2}&=&-\frac{1}{4}\Delta^2\zeta+{\mathcal L}(\chi,\zeta)- \hat{\gamma} \frac{\partial \zeta}{\partial t},\\
\Delta^2\chi&=&-\frac{1}{2}{\mathcal L}(\zeta,\zeta).
\end{eqnarray}
$\hat{\gamma}$ is the non-dimensional damping factor:

\begin{equation}
\hat{\gamma}=\gamma \tau=\gamma h \sqrt{\frac{\rho}{3E(1-\nu^2)}}.
\label{Eq34}
\end{equation}

In~\cite{humbert_ref7}, the damping law has been measured and behaves as $\gamma=\xi f^{0.6}$, where $\xi$ is a parameter taking different values, obtained by changing the configuration of the plate in a given manner. In order to use the same range of damping values in the phenomenological model as in the experiment, one has to express the relationship between the dimensional values of $\xi$ and their dimensionless counterparts $\hat\xi$. Thanks to Eq.~\eqref{Eq34}, we have
\begin{equation}
\hat{\gamma}= \tau(\xi f^{0.6})=\hat\xi\hat\omega^{0.6} \;\; {\rm with} \;\; \hat\xi=\frac{\tau^{0.4}}{(2 \pi)^{0.6}}\xi.
\end{equation}
Table 1 sums up the numerical $\xi$ values obtained from the experiments (first line, from \cite{humbert_ref15}) and their equivalent non-dimensional values $\hat\xi$ used previously for the simulations of the phenomenological model. Note that in the present paper and for the sake of simplicity, the coefficients used in the phenomenological model were named $\xi$.
\begin{table}[h!]
\label{tab.1}
\begin{center}
\begin{tabular}{lcccc}
$\xi$ & 0.045 & 0.072 & 0.14 & 0.22\\
$\hat\xi\times10^5$ & 1.908 & 3.0528 & 5.9359 & 9.3279
\end{tabular}
\caption{Correspondence between the experimentally measured values of damping coefficients $\xi$ and their dimensionless counterparts  $\hat\xi$ used in the simulations.}
\end{center}
\end{table}

\end{document}